\documentclass[sn-nature]{sn-jnl}
\usepackage{graphicx}
\usepackage{rotating}
\usepackage{tabularx}
\usepackage{longtable}
\usepackage{multirow}
\usepackage{amsmath,amssymb,amsfonts}
\usepackage{amsthm}
\usepackage{mathrsfs}
\usepackage[title]{appendix}
\usepackage{xcolor}
\usepackage{textcomp}
\usepackage{manyfoot}
\usepackage{booktabs}
\usepackage{makecell}
\usepackage{array}
\usepackage{placeins}
\usepackage[numbers]{natbib}

\usepackage{float}
\usepackage[normalem]{ulem}
\newfloat{extendedfigure}{htbp}{lof}
\floatname{extendedfigure}{Extended Data Fig.}
\newfloat{supfigure}{htbp}{lof}
\floatname{supfigure}{Supplementary Fig.}
\usepackage{caption}
\DeclareCaptionFont{small}{\small}
\DeclareCaptionLabelFormat{boldcolon}{\textbf{#1~#2}}

\newfloat{extendedtable}{htbp}{lot}  
\floatname{extendedtable}{Extended Data Table}

\newcommand{\araa}{Annu. Rev. Astron. Astrophys.} 
\newcommand{\aj}{Astron. J.} 
\newcommand{\apj}{Astrophys. J.} 
\newcommand{\apjl}{Astrophys. J. Lett.} 
\newcommand{\apjs}{Astrophys. J. Suppl. Ser.} 
\newcommand{\aap}{Astron. Astrophys.} 
\newcommand{\mnras}{Mon. Not. R. Astron. Soc.} 
\newcommand{\pasp}{Publ. Astron. Soc. Pac.} 
\newcommand{\ssr}{Space Sci. Rev.} 

\theoremstyle{thmstyleone}

\theoremstyle{thmstyletwo}%

\theoremstyle{thmstylethree}%

\begin{document}

\title[Bound from Birth]{The bound origin of low-mass stellar binaries}

\author*[1, 2, 3]{\fnm{Aleksey} \sur{Generozov}}\email{aleksey.generozov@gmail.com}

\author[2, 3]{\fnm{Stella S. R.} \sur{Offner}}\email{soffner@utexas.edu}
\equalcont{These authors contributed equally to this work.}

\author[4]{\fnm{Kaitlin M.} \sur{Kratter}}\email{kkratter@arizona.edu}
\equalcont{These authors contributed equally to this work.}
\author[1,5]{\fnm{Hagai B.} \sur{Perets}}\email{hperets@physics.technion.ac.il}
 \equalcont{These authors contributed equally to this work.}

\author[2]{\fnm{D{\'a}vid} \sur{Guszejnov}}\email{guszejnov.david@gmail.com}

\author[6]{\fnm{Michael Y.} \sur{Grudi{\'c}}}\email{mgrudic@flatironinstitute.org}

\affil[1]{\orgdiv{Physics Dept}, \orgname{Technion Israel Institute of Technology}, \orgaddress{\city{Haifa}, \postcode{32000}, \country{Israel}}}

\affil[2]{\orgdiv{Astronomy Dept}, \orgname{University of Texas at Austin}, \orgaddress{\city{Austin}, \state{TX} \postcode{78712}, \country{USA}}}

\affil[3]{\orgdiv{Oden Institute}, \orgname{University of Texas at Austin}, \orgaddress{\city{Austin}, \state{TX} \postcode{78712}, \country{USA}}}

\affil[4]{\orgdiv{Steward Obs and Astronomy Dept}, \orgname{University of Arizona}, \orgaddress{\city{Tuscon}, \state{AZ} \postcode{85721}, \country{USA}}}

\affil[5]{\orgdiv{Department of Natural Sciences}, \orgname{Astrophysics Research Center of the Open University (ARCO)}, \orgaddress{\city{Raanana}, \postcode{4353701}, \country{Israel}}}


\affil[6]{\orgdiv{Center for Computational Astrophysics}, \orgname{Flatiron Institute}, \orgaddress{\city{New York}, \state{NY} \postcode{10010}, \country{USA}}}

\abstract{Most main sequence stars, unlike our Sun, belong to multiple systems with two or more stars. How and when these multiples come together and become bound is uncertain, since the earliest stages of star formation are difficult to resolve. We analyze simulations of star cluster formation in Milky Way-like conditions, including all key physics and stellar feedback mechanisms, to understand how multiple systems form. We show that $\approx 70-80\%$ of binaries are gravitationally bound from the moment the second star forms. Binaries evolve and accrete together, which will affect their planetary systems and chemical evolution. Half of the binaries are disrupted by the end of the star-formation epoch, such that $\approx40\%$ of the final single stars belonged to a multiple at some point, with implications for the stellar initial mass function. Formation in multiples is the dominant mode of star formation, accounting for at least 57\% of stars. 
}

\maketitle

\section{Main}\label{sec:intro}
Observational studies of star formation have matured substantially over the past two decades due to comprehensive, multi-wavelength, high-resolution, surveys of nearby star-forming regions \citep{DunhamStutz2014a,PinedaArzoumanian2023a}. Material collapses from dense cores in molecular clouds. The cores are embedded in filamentary structure, generated by supersonic turbulence, self-gravity, and magnetic fields. When stars form, a majority become binary or higher order multiple systems rather than single stars \citep{OffnerMoe2023a}. The parent molecular cloud is dispersed on timescales of order 10 Myr, leaving a young star cluster \cite{chevance+2020}. Eventually the formed multiple systems are injected into the field after the birth cluster is disrupted (`infant mortality'; \cite{bastian+2011}).

Unlike the initial mass function, which depends weakly on the star-forming environment \citep{OffnerClark2014a, GuszejnovGrudic2022a}, the frequency and properties of binaries vary both with time and between clusters \citep{OffnerMoe2023a,GuszejnovRaju2023a}. Thus, binarity can give insight into the star forming environment and process. More broadly, understanding binary formation is crucial for modeling many phenomena that rely on binary properties, including stellar evolution,  reionization, stellar and compact-object mergers, supernovae, and planet formation \citep{Ibe+84,Li+98,Por+00, San+12,ConroyKratter2012a,Duc+13,Rafikov2013a,moe&kratter2021}. As binaries and multiples represent the dominant mode of star formation, understanding binary star formation is necessary for understanding the majority of star systems.

Understanding binary formation is not possible from observations alone, as early stages of fragmentation are inaccessible \cite{schnee+2010}. Instead, simulations of star cluster formation can be used to probe the formation of stellar multiples. Several simulations have shown good agreement between multiplicity statistics at the end of the star-forming phase and observations \cite{Bate2000a, bate2012,KrumholzKlein2012a,LiKleinMcKee2018,LeeOffner2019a,guszejnov+2023}, though the mechanisms governing binary formation remain relatively unexplored in the context of large statistical samples or comprehensive models that include all relevant physics (but see \cite{KuruwitaHaugbolle2023a}).

The most prominent wide binary formation mechanisms are typically described as either ``prompt fragmentation" or ``turbulent fragmentation" of dense gas cores \citep{Hoyle1953a,TsuribeInutsuka1999a,BossFisher2000a,PadoanNordlund2002a,Fisher2002a}. 
However, in contrast to simple analytical descriptions of star formation, cores are neither symmetric nor in solid body rotation \citep[e.g.,][]{PinedaArzoumanian2023a}. In the turbulent model, binary formation occurs when the non-linear perturbations expected in a turbulent cloud cause a sub-region to become over-dense and collapse more rapidly than the free-fall timescale of the background gas, thereby leading to the production of a secondary condensation. In such a model, the two perturbations need not be bound ab initio. The two stars form with initial separations of several thousand AU, where binding energies are small and easily comparable to the tidal field of the background cluster. 

A variety of work has shown that turbulent fragmentation is a viable mechanism for binary formation in simulations of star cluster formation \citep{OffnerKratter2010a,LeeOffner2019a,kuffmeier&calcutt2019,GuszejnovRaju2023a}. However, it is not yet clear whether binaries are born bound, or if they become captured later on via, e.g. gas dynamical friction (GDF), which is an efficient mechanism for subsequent orbital migration \citep{LeeOffner2019a}. In fact, recent analytic work suggested that 
GDF can dominate the binary production process in embedded clusters and other gas-rich environments \cite{RoznerGenerozov2023a}. The timing and mechanism for binary formation can significantly influence the dynamical evolution of binaries \cite{cournoyer-cloutier+2021}.
 
Here we analyze binary formation in {\sc starforge} simulations \citep{GrudicGuszejnov2022a,GuszejnovGrudic2022a}.
These are state-of-the-art ideal magnetohydrodynamics (MHD) simulations of star-forming clouds that include radiative cooling, thermochemistry, and feedback physics (see \S~\ref{sec:meth} for details). 

We now briefly summarize the initial conditions (described in detail in \S~\ref{sec:meth} and in \cite{GuszejnovGrudic2022a}) and the global evolution of the {\sc{Starforge}} simulations. The {\sc{starforge}} simulation suite includes clouds of different sizes, magnetic field strengths, and virial parameters. We focus our analysis on a fiducial cloud with properties typical of a Milky Way GMC, the \textbf{M2e4\_mu1.3} simulations from \cite{guszejnov+2022a} (see their Table 1). The initial mass and radius are 20,000 M$_\odot$ and 10 pc respectively, matching the mean surface density of GMCs in the Solar neighborhood (e.g. \cite{lada&dame2020}). The ratio of twice the kinetic to gravitational energy is $\alpha_{\rm vir}=2$, comparable to that of massive ($\gtrsim 10^4 M_{\odot}$) Milky Way GMCs \cite{larson1981,chevance+2023}. The cloud is initially threaded by a uniform magnetic field with $B_z \approx 6~\mu$G, corresponding to a mass-to-flux ratio, $\mu$, of 1.3. This field is comparable to that found by Zeeman surveys of molecular clouds \cite{crutcher+2010}. We analyze three realizations of such clouds with different initial turbulent seeds. 

As described in \cite{grudic+2021} the initial turbulence dissipates within a few crossing times, driving a drop in the virial parameter, along with localized collapse and star formation. Stars form in dense cores within filamentary structure and inject feedback in the form of radiation, protostellar outflows, stellar winds, and eventually, supernovae produced by massive stars. The star formation rate peaks, and then falls off as feedback drives out gas. The end result is that stars form clustered substructures, similar to observed complexes of young stars \cite{kuhn+2014,gouliermis2018,ward+2020}.

To test the robustness of our results, we also analyze clouds with initial magnetic fields $B_z \approx 2~\mu$G and $B_z \approx 20~\mu$G ($\mu = 4.2$ and 0.42 respectively), clouds with $\alpha_{\rm vir}=1$ and 4, and one cloud with subsolar metallicity ($Z/Z_{\odot}=0.1$). Finally, we analyze a simulation with fiducial parameters from a newer set of {\sc{starforge}} simulations with improved dust physics and radiative transfer (see \S~\ref{sec:meth} for details). This simulation is {\sc{starforge}} v1.2, while the others are v1.1. The number of realizations varies with simulation parameters: each magnetic field strength has three realizations and each virial parameters has two. In the other cases (v1.2 and low metallicity) there is only one simulation included in our analysis, as there is only one that has been run for at least 5 Myr. Altogether the fifteen simulations we analyzed formed 32526 stars, with 4425 multiples at the final simulation snapshots.

We note that none of the simulations we analyze resolve disks around the protostars, and thus they do not include multiples from disc fragmentation.

\section{Results}
Our work builds on previous analyses of multiple stars in the {\sc starforge} simulations. In particular, we use the algorithm described in  \cite{GuszejnovRaju2023a} for identifying multiple systems, with one addition: we include the mass of surrounding gas halos in the binding energy of young binaries. Due to the high resolution of the {\sc{starforge}} simulations, the sinks themselves are initially small and insignificant, and their potential is easily dominated by that of surrounding gas.

Briefly, to identify multiples we complete the following steps
(see \S~\ref{sec:meth} for details):
\begin{enumerate}
    \item Compute the gas halo masses around each star at every snapshot, and add the halo mass to each star's mass.
    \item Compute the binding energy of all nearby pairs of stars.
    \item Group the most bound pair into a system. This multiple is replaced with a single particle at the center-of-mass with the same total mass and momentum for all future binding energy calculations.
    \item Repeat steps 2 and 3 until no further multiples with multiplicity $\leq$4 are identified. 
\end{enumerate}

Note that unlike \cite{GuszejnovRaju2023a}, we do not impose any minimum mass ratio in our multiple identification.
Overall, we identify 1895 binaries that survive for at least one binary orbital period among the 6405 stars formed in the three fiducial simulations (see \S~\ref{sec:meth} for details). Each simulation is run for $\sim$10$^7$ yr, with output snapshots spaced by $24.7$ kyr. At the final snapshot of the simulations, there are 601 (isolated) binaries. Our overall multiple statistics are similar to those reported in \cite{GuszejnovRaju2023a}. For example, the multiplicity fraction

\begin{equation}
{\rm MF} = (B + T + Q) / (S + B + T + Q),
\end{equation}
where $S$, $B$, $T$, and $Q$ are the numbers of singles, binaries, triples, and quadruples for binaries with solar-type primaries ($0.7$--$1.3 M_{\odot}$) is $\sim$40\% at the end of the simulations. This is the multiplicity fraction excluding any cuts on lifetime (see \S~\ref{sec:binID}). Very roughly (within a factor of a few), this is comparable to the results of the semi-analytic core fragmentation model from \cite{guszejnov+2017} (see \cite{guszejnov+2023} for a detailed discussion).  

\begin{figure}[htbp!]
    \centering
    \includegraphics[height=0.43\textwidth]{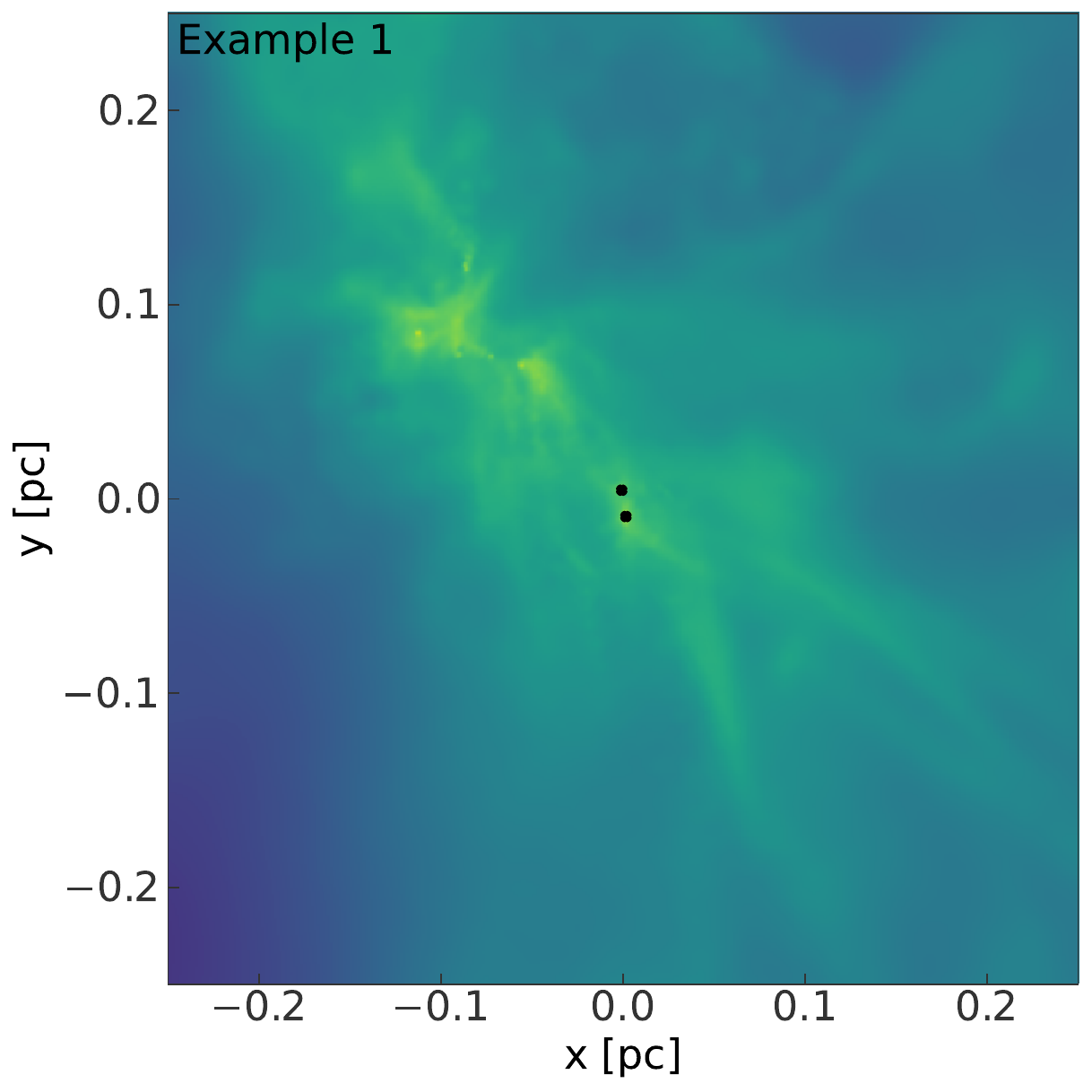}
    \includegraphics[height=0.43\textwidth]{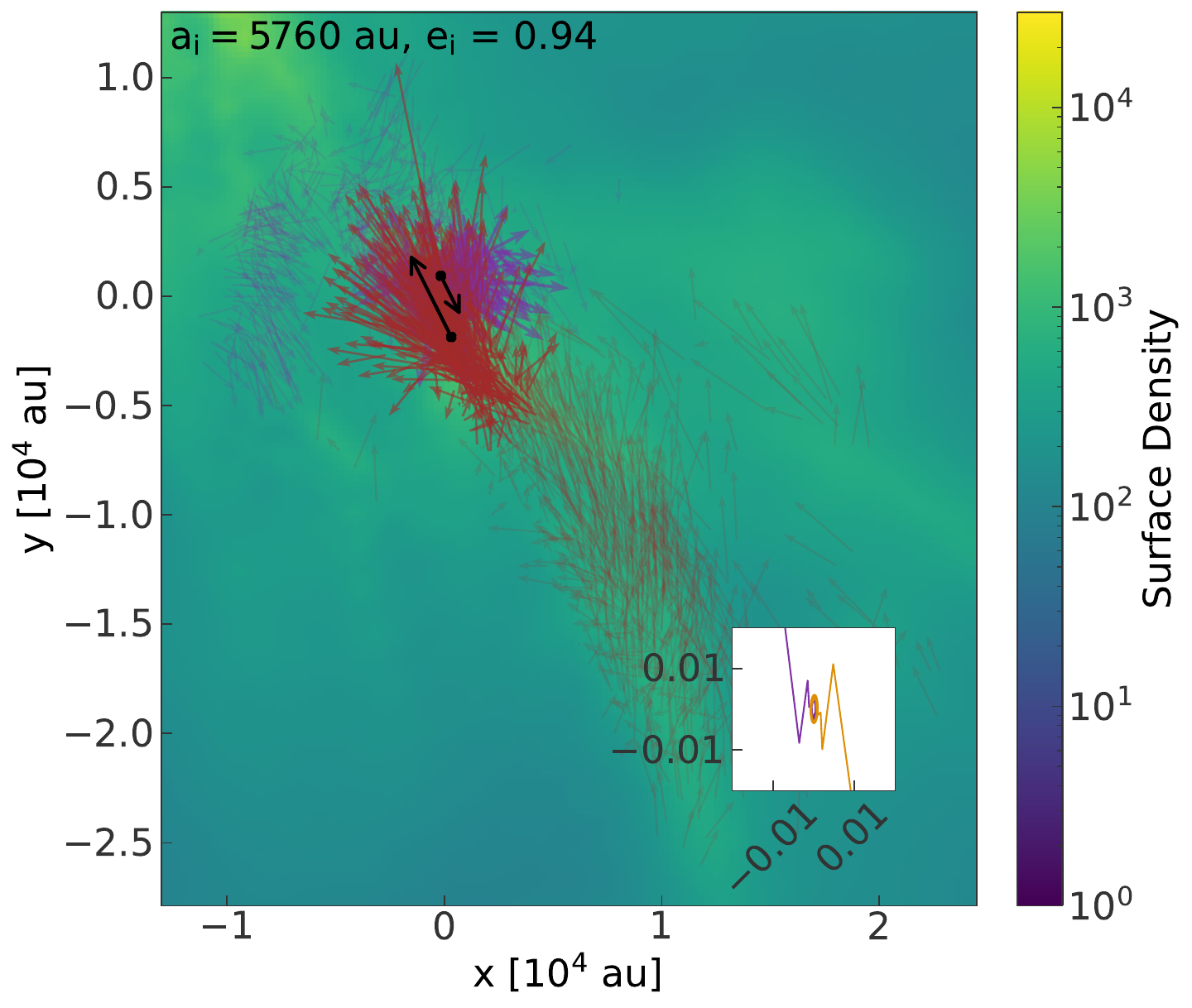}
    \includegraphics[height=0.43\textwidth]{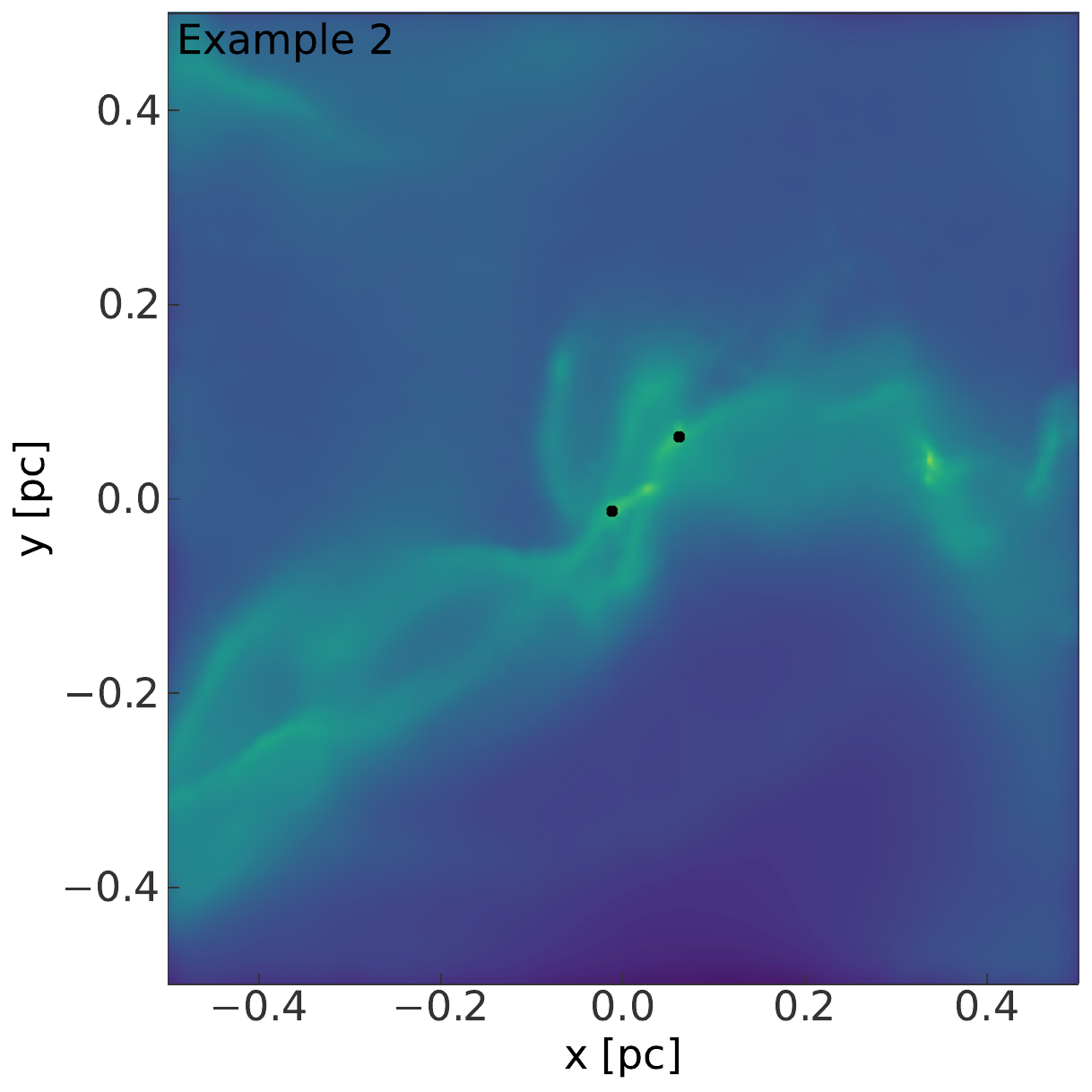}
    \includegraphics[height=0.43\textwidth]{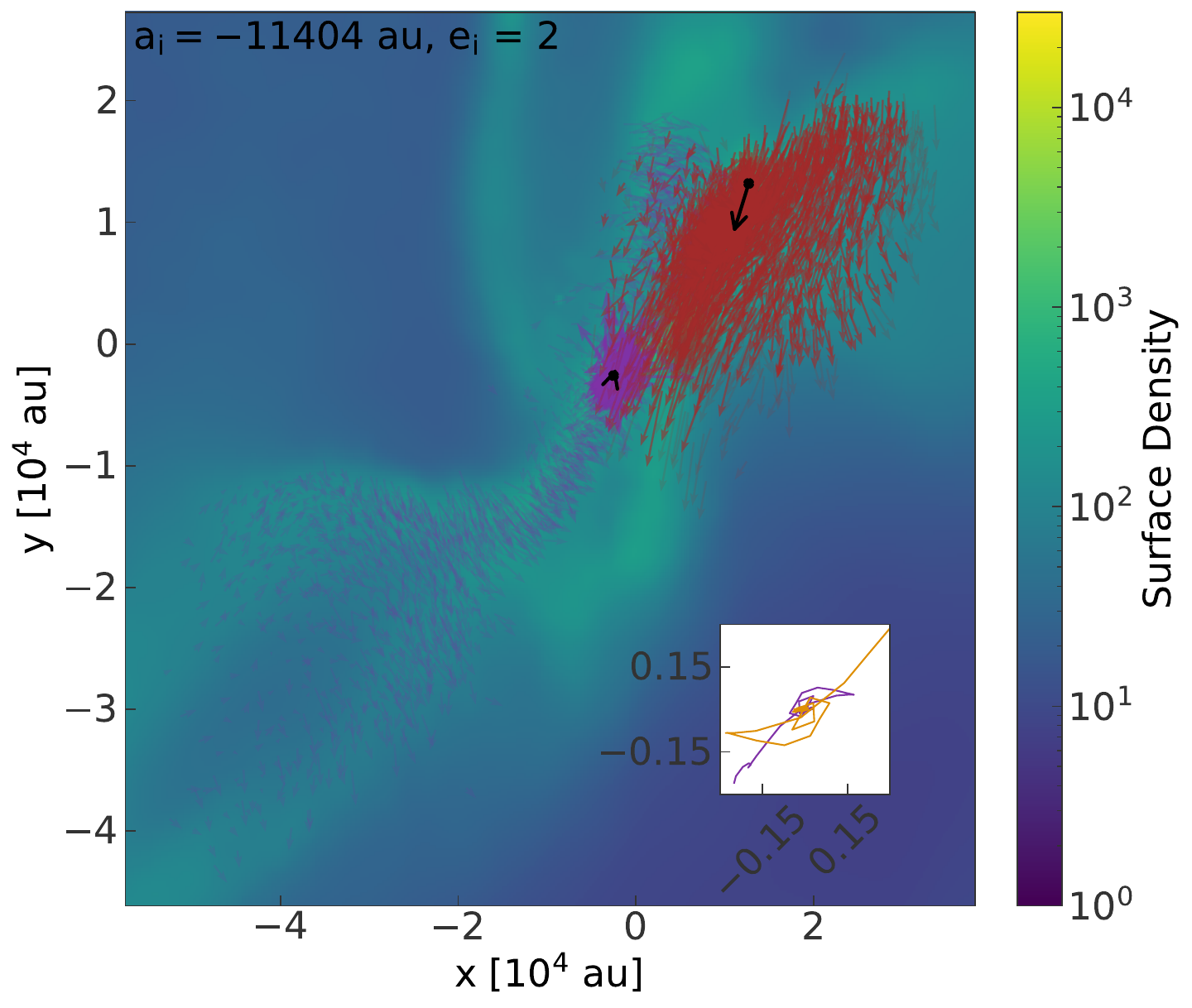}
    \caption{Example binaries and surrounding gas at formation. Each row shows an example binary and the surrounding gas surface density (along a 1 pc sightline) at the `Initial Snapshot Together' (IST). The top binary is Bound From Birth (BFB), and is only identified as such when the gas halo masses are included. The bottom binary is unbound at the IST, even after the inclusion of the gas. The right panels show a zoomed-in view and the velocities of identified gas halo particles with small, colored arrows. There is uncertainty in identifying the gas halo. The opaque arrows indicate our smaller, fiducial estimate, while the translucent arrows show our maximum estimate (see the discussion in Methods). The black arrows are the stars' velocity vectors at the IST.
    The arrow lengths are the distance particles would travel over two snapshots (49.4 kyr), if they traveled at constant velocity. All velocities are in the stars' center-of-mass frame. The orange and purple lines in the insets are the stars' center-of-mass trajectories after inspiral, with the same units as in the outer frame.}\label{fig:pretty}
\end{figure}

Initially, protostars are low mass and surrounded by a large gas halo, as in Figure~\ref{fig:pretty}. This figure shows two example binaries at the Initial Snapshot Together (IST)  - the first snapshot in which both stars exist. Binaries typically form with a separation of $10^3-10^4$ au, and then inspiral and circularize, as in Example 1 in Figure~\ref{fig:pretty} (see also Extended Data Fig.~\ref{fig:evolve}). The inspiral is gas-driven, and proceeds until the gas halos are depleted. In some cases, like Example 2, the binary is significantly perturbed by interactions with other stars: as shown in Extended Data Fig.~\ref{fig:evolve}, the binary stars are bound to other partners at some points of the inspiral.

\begin{figure}[h]
    \centering
    \includegraphics[width=0.45\textwidth]{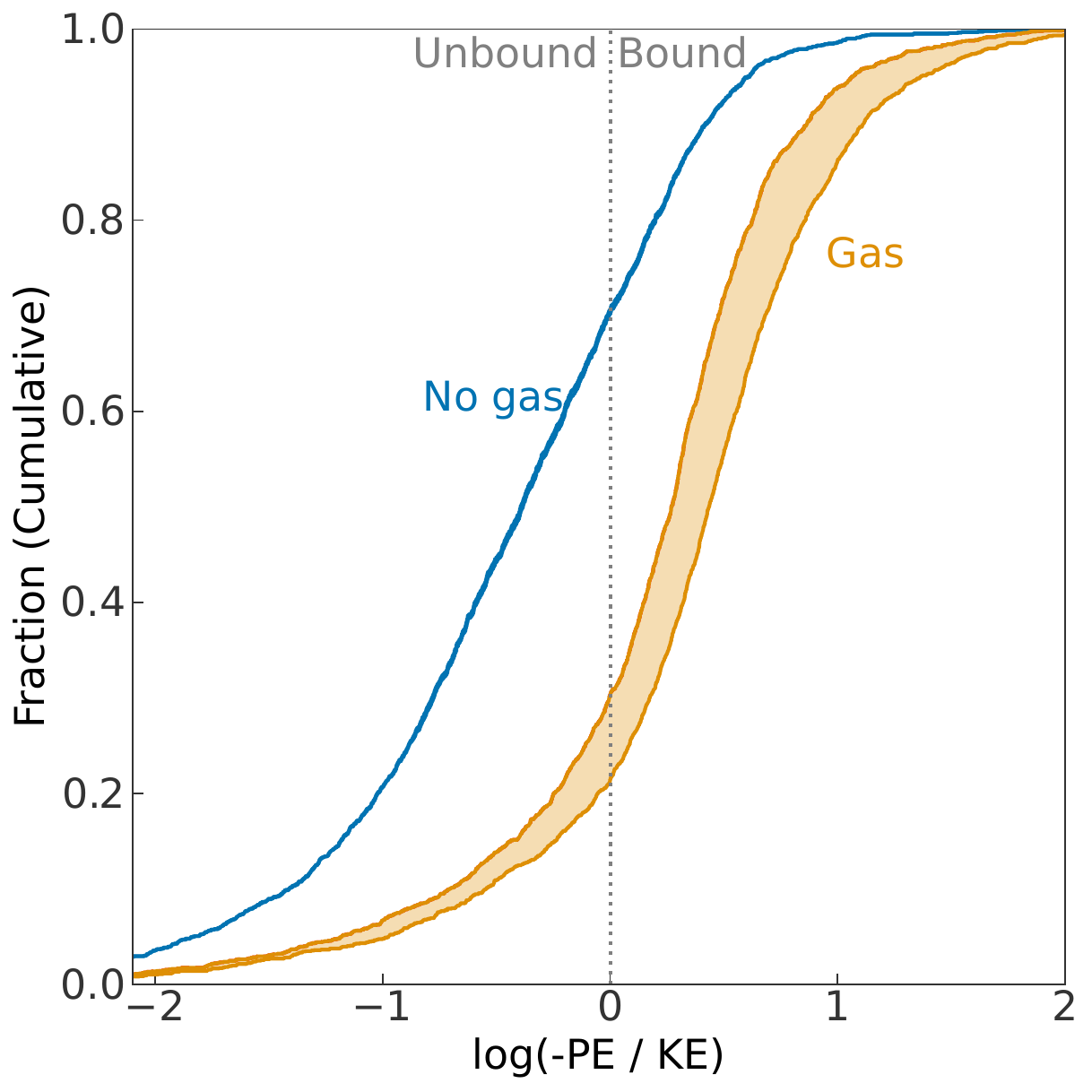}
    \includegraphics[width=0.45\textwidth]{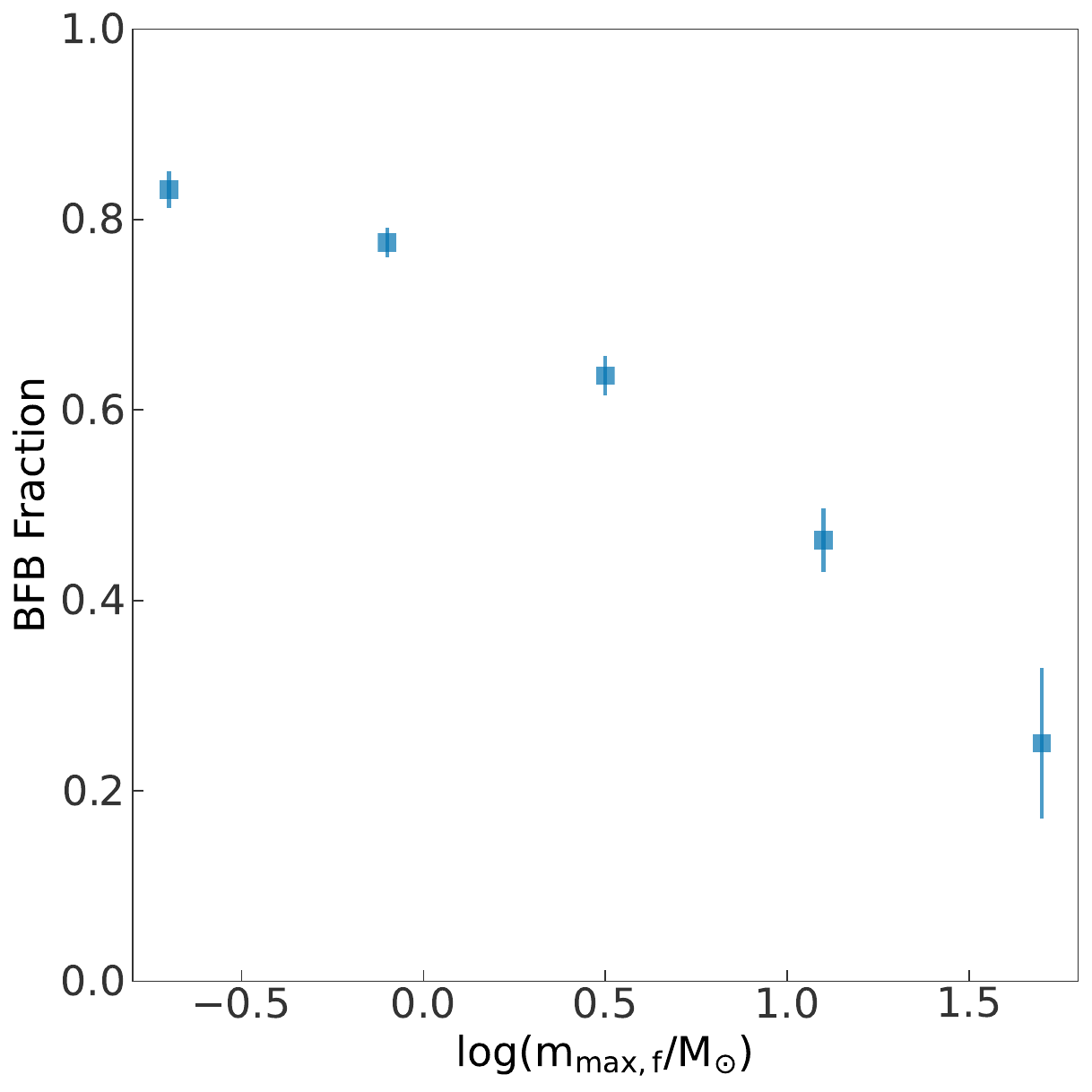}
    \caption{Boundedness of binary pairs at formation. Left panel: Binding energy of all binary pairs at the IST, normalized to the kinetic energy (in the center of mass frame).  Generally, when this ratio is greater than one, the binary is bound from birth. Conversely, when the ratio is less than one there is a delay between star and binary formation. While there are exceptions due to higher order multiples, they have a negligible impact on this distribution. The number of pairs in the unbound group is significantly reduced if the gas halo mass is included in the binary binding energy, as shown by the difference between the blue (no gas) and orange (gas) histograms. The shaded regions show the uncertainty due to the calculation of the gas halo (see Methods for details). Right panel: Fraction of binaries that are bound from birth as a function of the maximum of the two stars' final masses. The error bars are calculated using the Bayesian formula $\sqrt{\frac{(N - k + 1) (k + 1)}{(N+ 3) (N + 2)^2}}$ \cite{guszejnov+2023}, where $k$ is the number of bound binaries and $N$ is the total number of binaries in the mass bin. From left to right $k$ is 311, 565, 339, 102, 7, while $N$ is 374, 728, 533, 220, 28.}
    \label{fig:energy}
\end{figure}

The left panel of Fig.~\ref{fig:energy} shows the binding energy of binaries at the IST. 
 Most pairs ($\sim$70\%)  appear gravitationally unbound when the potential of the surrounding gas halos is neglected. However, the unbound fraction drops to $\sim$21--30\%, so that $\sim$70--80\% of pairs are bound in the same structure, when the potential of the gas halo is included. For this calculation, we do not impose any constraints on the lifetime of these structures. In general, we reserve the term `multiple system' for configurations that complete at least one orbit. Here the ranges reflect the uncertainty in the halo mass calculation (see the discussion in \S~\ref{sec:gas}).  While most systems are bound from birth, like Example 1 in Fig.~\ref{fig:pretty}, a significant minority ($\sim$21--30\%) are not, even after the inclusion of the gas potential. Approximately 60\%--69\% of the unbound binaries ($\sim$15-18\% of the total) had one star in a multiple system before the two stars became bound, indicating an exchange origin. An exchange origin for these binaries is also supported by the mass dependence of the bound from birth fraction, which ranges from $\sim$80\% at 0.2 $M_{\odot}$ to $\sim$25\% at $50 M_{\odot}$ (see the right panel of Fig.~\ref{fig:energy}). This mass dependence is expected, because massive stars form in denser environments and are more likely to experience dynamical interactions. While massive stars tend to form in multiples (Extended Data Fig.~\ref{fig:delayMult}), many are disrupted by exchange interactions. Thus, massive stars are born in multiples, and retain companions, but undergo substantial dynamical exchanges that rarely leave primordial systems intact  (Extended Data Fig.~\ref{fig:bfbMass}).

\begin{figure}[h]
    \centering
    \includegraphics[width=0.49\textwidth]{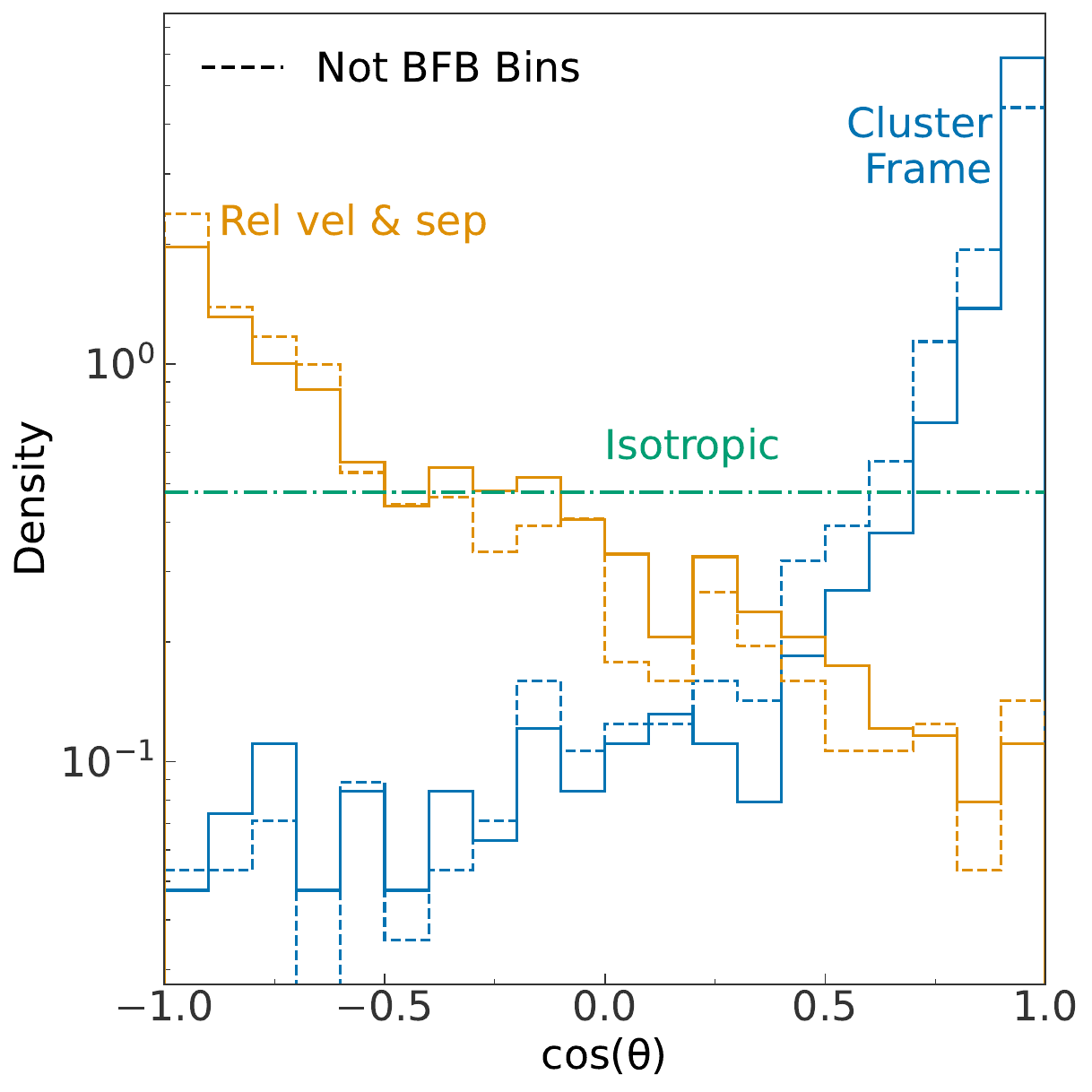}
    \includegraphics[width=0.49\textwidth]{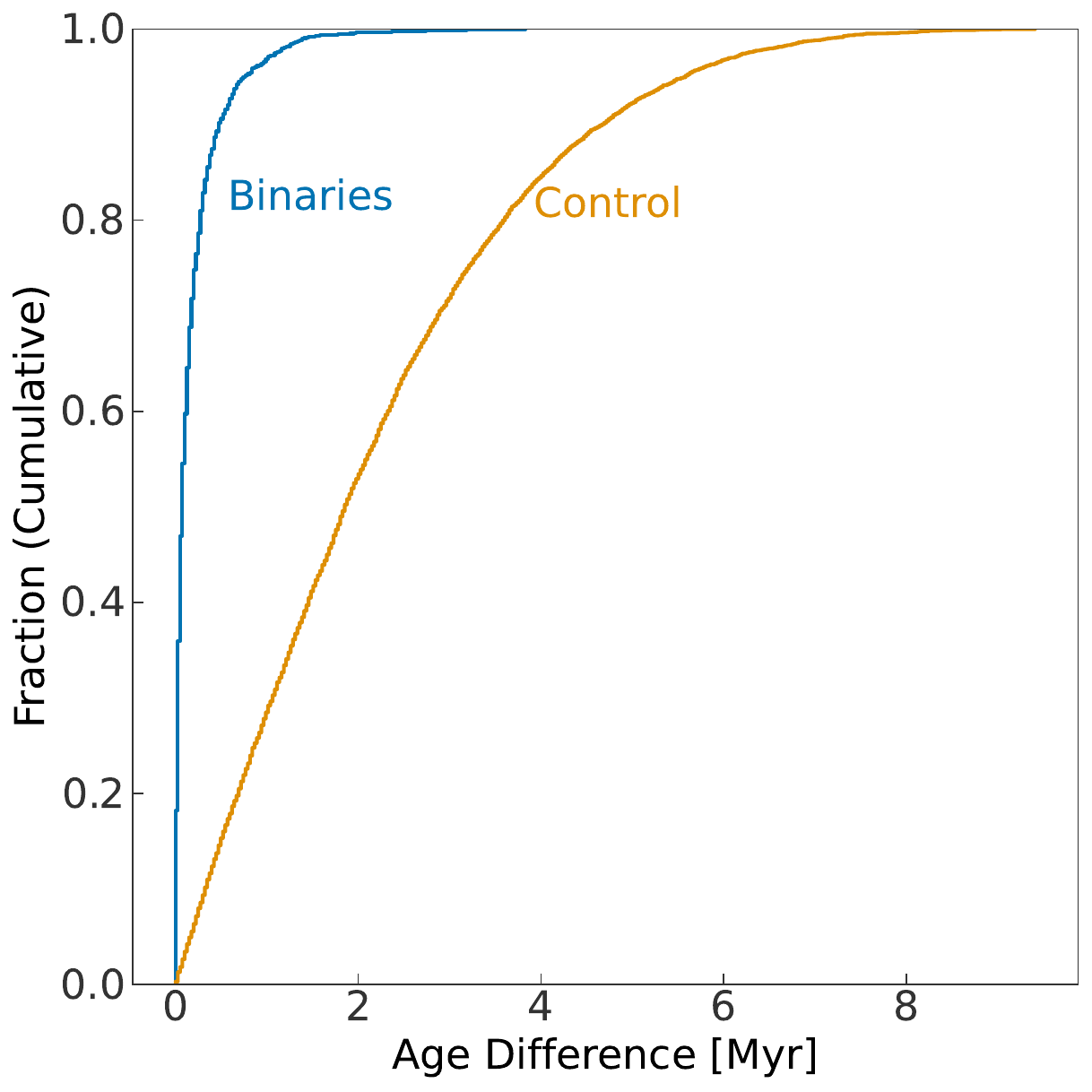}
    \caption{Distribution of binary velocities and formation times. Left panel: The blue, solid histogram shows the distribution of the angle between the stellar velocities (in the cluster frame) for each binary at the IST. The binary stars' velocities are initially highly aligned, and far from isotropic (green, dash-dotted line). The orange, solid histogram shows the angle between the relative velocity and the separation. An excess near $\cos(\theta)=-1$ corresponds to radially in-falling trajectories. Remarkably, the distributions for initially unbound binaries are similar, as shown by the dashed histograms. Right panel: Cumulative distribution of age differences between binary stars (blue) compared to the age differences
    of randomly chosen pairs (orange). Most binary stars form within a few$\times 10^5$ yr of each other. Thus, their age differences are comparable to or less than the protostellar lifetime.}\label{fig:angAge}
\end{figure}

Binary stars tend to form with highly correlated velocities, even if they are initially unbound, as illustrated in  Fig.~\ref{fig:pretty} and Fig.~\ref{fig:angAge}. The left panel of Fig.~\ref{fig:angAge} shows the distribution of the angle between the velocities of binary stars at the IST. In the cluster frame, the stellar velocities are nearly aligned. In each star's frame, the PDF shows an excess of radially infalling velocities. This correlation suggests that binary formation is mostly determined by the gas initial conditions, rather than by subsequent dynamical evolution and capture. In addition, binary stars typically form within a few$\times 10^5$ yr of each other (right panel of Fig.~\ref{fig:angAge}; see also Fig.~11 in \citealp{LeeOffner2019a}). This timescale is comparable to the free-fall time of a gas core and within the estimated protostellar lifetime, during which stars accrete nearly all of their mass \citep{DunhamStutz2014a,kristensen&dunham2018}.

Table~\ref{tab:binStats} compares the fraction of binaries that are bound from birth in the fiducial simulations with that in two other simulations with different initial magnetic fields. Higher magnetization delays star formation and suppresses the formation of high mass stars \cite{guszejnov+2022a}. The fraction of binaries bound from birth is between 70 and 73\% across all these simulations. Extended Data Table~\ref{tab:binStatsVir} shows that changing the virial parameter and the metallicity have a minor effect on the fraction of binaries that are bound from birth. These tables show the bound fractions for our smaller estimate of the halo mass. The bound fractions could be even higher considering the uncertainty in calculating the halo.

\begin{figure}[h]
    \centering
    \includegraphics[width=0.95\textwidth]{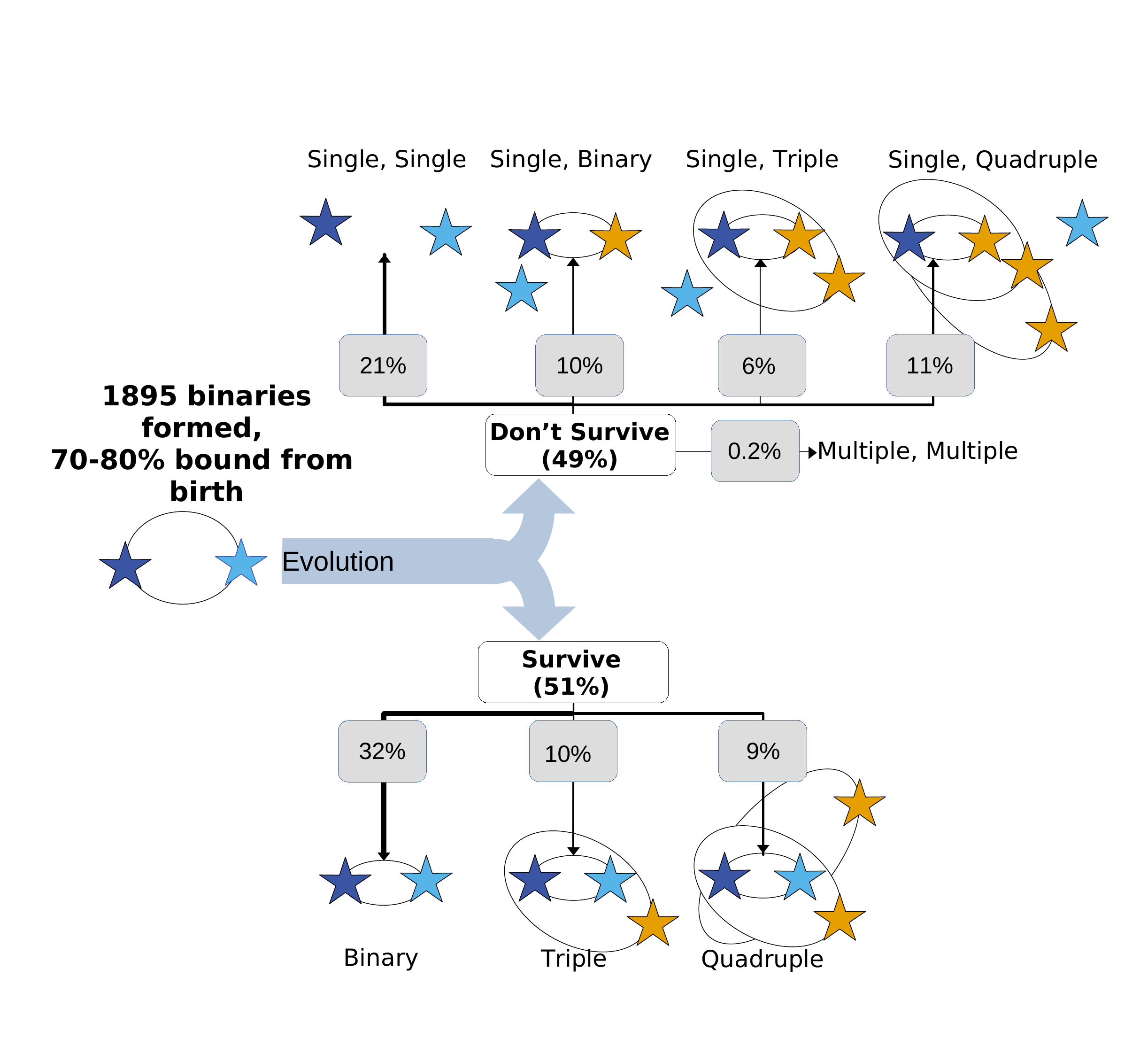}
    \caption{Schematic summarizing basic statistics on binary formation and survival. The bottom branches show the fates of the $\sim$50\% of binaries that survive, while the top branches show the end states of binaries that do not. The range in the bound from birth fraction, reflects the uncertainty in the halo mass estimate.}\label{fig:schema}
\end{figure}

Table~\ref{tab:binStats} also shows statistics for binaries surviving at the final snapshot, at which point most of the cloud gas is dispersed. We also include binaries with stars that go supernova (11 stars in total) among the surviving group, as long as they remain bound until the supernova. At the final time, the system configurations are largely fixed and representative of the expected main sequence multiplicity, which is consistent with observations, \citep{MoeKratter2018a}. Of the 1895 binaries formed in the fiducial simulations, 968 ($\sim$51\%) are in the same multiple system at the final snapshot (see Table~\ref{tab:binStats}).  Fig.~\ref{fig:schema} shows in detail the fates of these surviving binaries (e.g. the fraction that are in higher order multiples). Eventually the surviving binaries will become field multiples after the cluster disperses \cite{bastian+2011,farias+2024,farias+2025}.

The surviving binaries have a bimodal separation distribution with a large peak near 10 au and a small peak near 400 au. The location of the dominant peak is comparable to the peak of the semi-major axis distribution for evolved binaries from the star cluster simulations of \cite{cournoyer+2024} (see their Fig. 6). However, this comparison should be interpreted with caution. Binaries are injected directly in the simulations of \cite{cournoyer+2024}, with an observationally-motivated distribution. \cite{moe&distefano2017,winters+2019,offner+2023}. However, this distribution may already be evolved rather than primordial \cite{tokovinin&moe2020}. In our case, the separation distribution is artificially affected by gravitational softening.

\begin{table}[]
\caption{Summary of binary statistics for three different initial magnetic fields ($B_z \approx 6~\mu$G, $B_z \approx 2~\mu$G, and $B_z \approx 20~\mu$G, corresponding to mass-to-flux ratios, $\mu$, of 1.3, 4.2, and 0.42). The last two columns show the statistics from a simulation with updated dust and radiation physics (with $\mu=1.3$). The first two rows show the total number of binaries and the number and fraction bound at the IST. The next two rows show the same data for binaries that survive to the final snapshot (including those that are part of higher order multiples). The first two columns correspond to our fiducial simulations. We show data for individual simulations with different initial random seeds on the bottom of each cell, and stacked data at the top in bold. With the exception of v1.2, we have three seeds for each simulation. All the statistics correspond to the smaller of our two estimates of the halo mass (see Methods).  \label{tab:binStats}}
\begin{tabular}{|l|c|c|c|c|c|c|c|c|}
\hline
 & \multicolumn{2}{c|}{\texttt{mu1.3}} & \multicolumn{2}{c|}{\texttt{mu4.2}} & \multicolumn{2}{c|}{\texttt{mu0.4}} & \multicolumn{2}{c|}{\texttt{mu1.3, v1.2}} \\
 & \multicolumn{2}{c|}{seeds=3} & \multicolumn{2}{c|}{seeds=3} & \multicolumn{2}{c|}{seeds=3} & \multicolumn{2}{c|}{seeds=1} \\
\hline
 & Count & Frac & Count & Frac & Count & Frac & Count & Frac \\
\hline
\multicolumn{9}{|c|}{\textbf{All}}\\
\hline
Total &
\begin{tabular}{@{}c@{\hskip 1pt}c@{\hskip 1pt}c@{}}
 & \textbf{1895} & \\
{\footnotesize 581} & {\footnotesize 639} & {\footnotesize 675} 
\end{tabular} & 
\begin{tabular}{@{}c@{\hskip 1pt}c@{\hskip 1pt}c@{}}
 & 1.00 & \\
 & & 
\end{tabular} 
&
\begin{tabular}{@{}c@{\hskip 1pt}c@{\hskip 1pt}c@{}}
 & \textbf{1792} & \\
{\footnotesize 545} & {\footnotesize 600} & {\footnotesize 647}
\end{tabular} & 
\begin{tabular}{@{}c@{\hskip 1pt}c@{\hskip 1pt}c@{}}
 & 1.00 & \\
 & & 
\end{tabular} 
&
\begin{tabular}{@{}c@{\hskip 1pt}c@{\hskip 1pt}c@{}}
 & \textbf{2867} & \\
{\footnotesize 1115} & {\footnotesize 1015} & {\footnotesize 737}
\end{tabular} 
&
\begin{tabular}{@{}c@{\hskip 1pt}c@{\hskip 1pt}c@{}}
 & 1.00 & \\
 & & 
\end{tabular} 
&
\begin{tabular}{@{}c@{\hskip 1pt}c@{\hskip 1pt}c@{}}
 & \textbf{449} & \\
 & &
\end{tabular} 
& 
\begin{tabular}{@{}c@{\hskip 1pt}c@{\hskip 1pt}c@{}}
 & 1.00 & \\
 & & 
\end{tabular} 
\\
\hline
Bound &
\begin{tabular}{@{}c@{\hskip 1pt}c@{\hskip 1pt}c@{}}
 & \textbf{1332} & \\
{\footnotesize 403} & {\footnotesize 446} & {\footnotesize 483}
\end{tabular} 
& 
\begin{tabular}{@{}c@{\hskip 1pt}c@{\hskip 1pt}c@{}}
 & \textbf{0.70} & \\
{\footnotesize 0.69} & {\footnotesize 0.70} & {\footnotesize 0.72}
\end{tabular} 
&
\begin{tabular}{@{}c@{\hskip 1pt}c@{\hskip 1pt}c@{}}
 & \textbf{1301} & \\
{\footnotesize 405} & {\footnotesize 436} & {\footnotesize 460}
\end{tabular} 
&
\begin{tabular}{@{}c@{\hskip 1pt}c@{\hskip 1pt}c@{}}
 & \textbf{0.73} & \\
{\footnotesize 0.74} & {\footnotesize 0.73} & {\footnotesize 0.71}
\end{tabular} 
&
\begin{tabular}{@{}c@{\hskip 1pt}c@{\hskip 1pt}c@{}}
 & \textbf{2058} & \\
{\footnotesize 796} & {\footnotesize 725} & {\footnotesize 537}
\end{tabular} 
&
\begin{tabular}{@{}c@{\hskip 1pt}c@{\hskip 1pt}c@{}}
 & \textbf{0.72} & \\
{\footnotesize 0.71} & {\footnotesize 0.71} & {\footnotesize 0.73}
\end{tabular} &
\begin{tabular}{@{}c@{\hskip 1pt}c@{\hskip 1pt}c@{}}
 & \textbf{325} & \\
 & &
\end{tabular} 
& 
\begin{tabular}{@{}c@{\hskip 1pt}c@{\hskip 1pt}c@{}}
 & \textbf{0.72} & \\
 & & 
\end{tabular} \\
\hline
\multicolumn{9}{|c|}{\textbf{Surviving}}\\
\hline
Total &
\begin{tabular}{@{}c@{\hskip 1pt}c@{\hskip 1pt}c@{}}
 & \textbf{968} & \\
{\footnotesize 292} & {\footnotesize 338} & {\footnotesize 338}
\end{tabular} 
& 
\begin{tabular}{@{}c@{\hskip 1pt}c@{\hskip 1pt}c@{}}
 & 1.00 & \\
 & & 
\end{tabular} 
&
\begin{tabular}{@{}c@{\hskip 1pt}c@{\hskip 1pt}c@{}}
 & \textbf{938} & \\
{\footnotesize 292} & {\footnotesize 318} & {\footnotesize 328}
\end{tabular} 
& 
\begin{tabular}{@{}c@{\hskip 1pt}c@{\hskip 1pt}c@{}}
 & 1.00 & \\
 & & 
\end{tabular} 
&
\begin{tabular}{@{}c@{\hskip 1pt}c@{\hskip 1pt}c@{}}
 & \textbf{1489} & \\
{\footnotesize 597} & {\footnotesize 512} & {\footnotesize 380}
\end{tabular} 
& 
\begin{tabular}{@{}c@{\hskip 1pt}c@{\hskip 1pt}c@{}}
 & 1.00 & \\
 & & 
\end{tabular} 
&
\begin{tabular}{@{}c@{\hskip 1pt}c@{\hskip 1pt}c@{}}
 & \textbf{207} & \\
 & &
\end{tabular} 
& 
\begin{tabular}{@{}c@{\hskip 1pt}c@{\hskip 1pt}c@{}}
 & 1.00 & \\
 & & 
\end{tabular} \\
\hline
Bound &
\begin{tabular}{@{}c@{\hskip 1pt}c@{\hskip 1pt}c@{}}
 & \textbf{682} & \\
{\footnotesize 197} & {\footnotesize 236} & {\footnotesize 249}
\end{tabular} 
& 
\begin{tabular}{@{}c@{\hskip 1pt}c@{\hskip 1pt}c@{}}
 & \textbf{0.70} & \\
{\footnotesize 0.67} & {\footnotesize 0.70} & {\footnotesize 0.74}
\end{tabular} 
&
\begin{tabular}{@{}c@{\hskip 1pt}c@{\hskip 1pt}c@{}}
 & \textbf{670} & \\
{\footnotesize 212} & {\footnotesize 220} & {\footnotesize 238}
\end{tabular} 
&
\begin{tabular}{@{}c@{\hskip 1pt}c@{\hskip 1pt}c@{}}
 & \textbf{0.71} & \\
{\footnotesize 0.73} & {\footnotesize 0.69} & {\footnotesize 0.73}
\end{tabular} 
&
\begin{tabular}{@{}c@{\hskip 1pt}c@{\hskip 1pt}c@{}}
 & \textbf{1069} & \\
{\footnotesize 423} & {\footnotesize 376} & {\footnotesize 270}
\end{tabular} 
&
\begin{tabular}{@{}c@{\hskip 1pt}c@{\hskip 1pt}c@{}}
 & \textbf{0.72} & \\
{\footnotesize 0.71} & {\footnotesize 0.73} & {\footnotesize 0.71}
\end{tabular} 
&
\begin{tabular}{@{}c@{\hskip 1pt}c@{\hskip 1pt}c@{}}
 & \textbf{145} & \\
 & &
\end{tabular} 
& 
\begin{tabular}{@{}c@{\hskip 1pt}c@{\hskip 1pt}c@{}}
 & \textbf{0.70} & \\
 & & 
\end{tabular} \\
\hline
\end{tabular}
\end{table}

\begin{figure}
    \centering
    \includegraphics[width=0.49\textwidth]{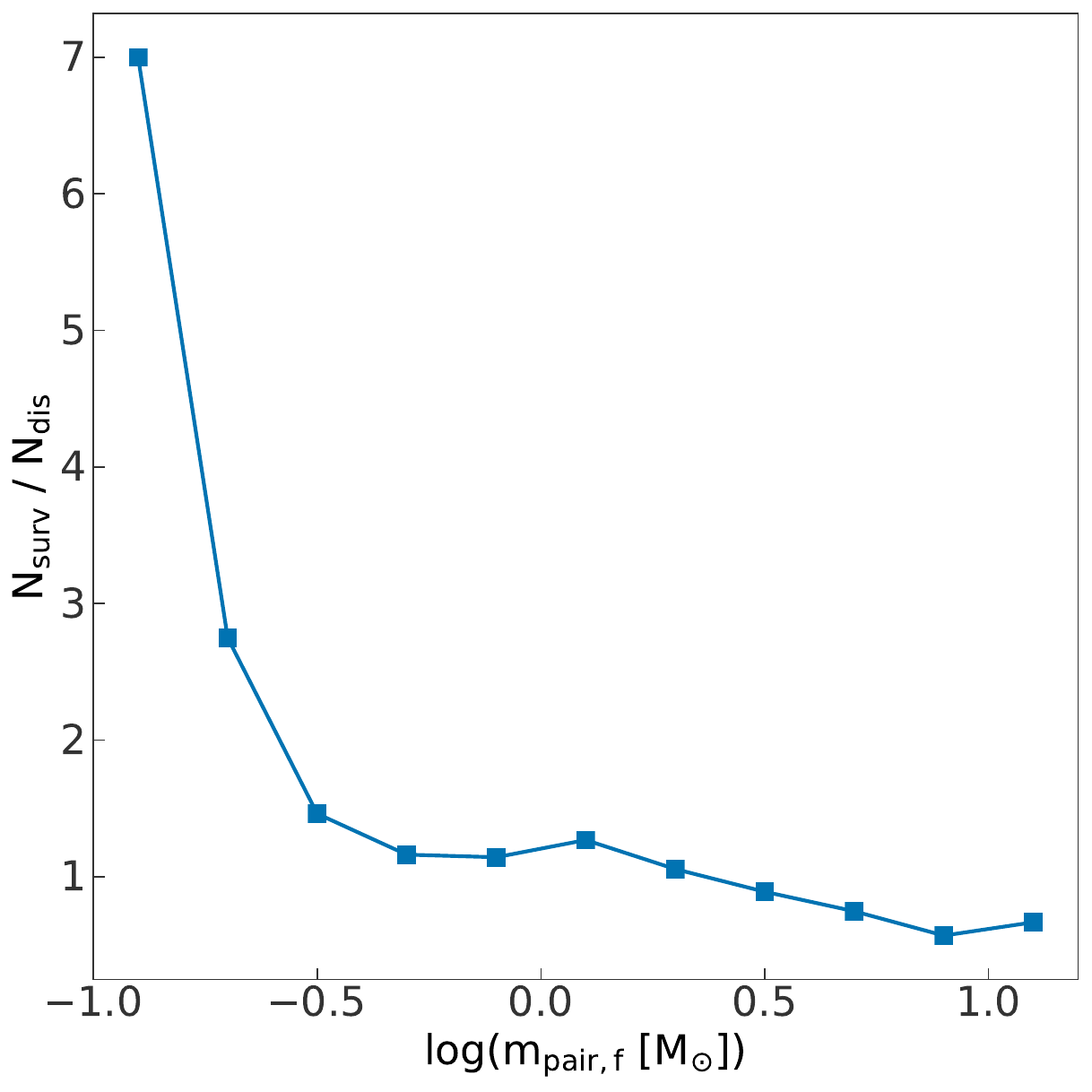}
    \caption{Ratio of the number of surviving to disrupted binaries versus the final mass of the binary stars. Here, surviving binaries are all those that are in the same system at the end. Disrupted binaries those that are in different multiples or ionized.}
    \label{fig:close}
\end{figure}

Fig.~\ref{fig:schema} also shows non-surviving binaries have different evolutionary paths: for $\sim$56\% of them at least one star is in a multiple at the end of the simulation. The remainder are single stars. Stars from the latter group (``ionized binaries'') are likely disrupted via close encounters and multiple interactions. 
Tracking close encounters directly is nontrivial, considering the limited time cadence of the simulations.
However, for approximately half of the ionized binaries, one of the stars is immediately (but temporarily) bound to other stars as a member of a multiple system after the disruption. 
Here we only count multiples that persist for at least one period. If we drop this requirement, then 69\% of ionized binaries have a star in a multiple after disruption.
The ratio of surviving to non-surviving binaries decreases with binary mass, as shown in Fig.~\ref{fig:close}. This decline is likely because massive stars are forming in denser environments and in higher order multiples, and are more likely to encounter other stars \citep{guszejnov+2023}.

Interestingly, the separation distribution of non-surviving binaries is not significantly different from the final separation distribution of surviving binaries. In both cases, $\sim80\%$ of pairs at some point had pericenters within 40 au, which is twice the typical gravitational softening length in the simulations. This suggests many hard binaries are disrupted, where hard binaries are those with semi-major axes below the hard-soft boundary
\begin{align}
    a_{\rm hs} &= \frac{G m_1 m_2}{2 \langle m \rangle \sigma^2} \nonumber\\
    & = 50\, {\rm au} \left(\frac{\sigma}{3\, {\rm km\, s^{-1}}}\right)^{-2} \left(\frac{m_1}{1 M_{\odot}}\right) \left(\frac{m_2}{1 M_{\odot}}\right) \left(\frac{\langle m \rangle}{1 M_{\odot}}\right)^{-1},
    \label{eq:hs}
\end{align}
where $\sigma$ is the 1D stellar velocity dispersion, $m_1$ and $m_2$ are the masses of the binary stars, and $\langle m \rangle$ is the mean mass of the surrounding stars \cite{heggie1975}. Quantitatively $\sim 70\% $ of the disrupted binaries are hard immediately prior to disruption (for comparison $\sim 80\%$ of the binaries at the end are hard). We compute $a_{\rm hs}$ for each binary using the local velocity dispersion and mean stellar mass estimated from the four nearest neighbors. Generally, hard binaries tend to become more bound due to interactions with surrounding stars and survive, while soft binaries tend to become less bound and disrupt \cite{heggie1975}. However, even hard binaries can be disrupted through exchange interactions, despite being energetically resistant to direct ionization. We also note the destruction rate of close binaries is affected by gravitational softening, though the sign of the effect is unclear. On the one hand, softening stalls binary inspiral, which would make binaries easier to disrupt. On the other hand, the softening could reduce close encounters between stars and reduce the likelihood of disruption.

So far, we have included binaries with stars below the completeness limit of $0.1 M_{\odot}$. Such binaries are 26\% of the total formed. However, we find excluding  binaries with a companion below 0.1 $M_{\odot}$ does not have a strong effect on the statistics in Fig.~\ref{fig:schema}.

Binary destruction can form a special population of single stars.
At the end of the simulation (after $10-12$ Myr), $\sim$26\% of all single stars were at one time members of binaries. This fraction is $\sim$59\% for stars $\geq 1 M_{\odot}$. The fraction that were in multiple systems (binaries, triple, or quadruples) is 41\% (67\% for stars $\geq 1 M_{\odot}$). We note that stars may have been part of a binary and a higher multiple either at the same time or at different times. 
Overall, a significant fraction of observed single stars may have a multiple origin that may affect their mass function.  

\section{Discussion and Summary}
Interpreting observations of binary stars and their planetary systems requires an understanding of how stellar binaries form. For example, the time at which stars acquire companions can affect the architecture of their planetary systems throughout time, considering that the early presence of a companion can truncate protoplanetary disks. Furthermore the simultaneous formation is a key, untested assumption of binary population synthesis models.

To gain insight into binary formation, we analyze state-of-the-art simulations of star forming clouds. Our findings suggest that most ($\sim$70--80\%) binaries form gravitationally bound from birth. This fraction depends on mass, ranging from $\sim$80\% at 0.2 $M_{\odot}$ to $\sim$25\% at 50 $M_{\odot}$ (see the right panel of Fig.~\ref{fig:energy} and Extended Data Fig.~\ref{fig:bfbMass}). While massive stars tend to form in multiples, many are disrupted by interactions. Many low mass stars, on the other hand, do not form in multiple systems (see Extended Data Fig.~\ref{fig:delayMult}). Nonetheless, a small majority (57\%) of all stars are in a multiple system within $5\times 10^5$ yr of formation. The true fraction of stars that form in multiples is likely even higher in nature, as the multiplicity fraction of low mass stars in the simulations is artificially reduced by spatial and mass resolution effects and the absence of disks.

Binary stars form at similar times (see Fig.~\ref{fig:angAge}), generally surrounded by large gas halos. These halos play a critical role in binding them at formation when the stars are small, but are quickly lost (generally within $\sim$1 Myr). Typically, the final mass of the star is $\sim$15--20\% of the maximum mass of each halo, though the distribution is extremely broad (Extended Data Fig.~\ref{fig:accEfficiency}). The maximum halo mass increases sublinearly with the final stellar mass, such that the ratio of the maximum halo mass to the final stellar mass is less than one for the most massive stars. 
This happens in part because the gas accreted by massive stars is never all contained in the identified halo at the same time. Instead, gas that is accreted flows into the circumstellar region over an extended period of time \cite{padoan+2020,grudic+2021}. Moreover, our halo identification is done star-by-star, whereas in reality, multiple stars may accrete from a shared halo—especially massive stars in dense, crowded environments. In fact, for stars $\gtrsim 10 M_{\odot}$ only $\sim 20\%$ of accreted gas particles are ever identified in their individual halos.

As binary stars form together, they will likely not have significant age or chemical differences. Furthermore, companions will affect the formation of planetary systems from the beginning, potentially truncating the outer portions of protoplanetary disks.

Early dynamics do affect binaries' \emph{fate} 
during the star-formation epoch, unbinding $\sim$50\% of them. Mostly, the non-surviving binaries are split either into a single star and a multiple, or two single stars by close encounters with other stars.

 Approximately forty percent of the final single-star population (4314 stars) belong to multiple systems at some point (67\% for stars $\geq 1 M_{\odot}$). This multiple origin may leave a statistical imprint on these stars' mass function. This is also motivated by observations,  considering the observed mass functions of binary primaries and secondaries differ \citep{MoeDi-Stefano2017a}.  However, this multiple imprint is modest in our simulations. The overall IMF of single stars closely resembles that of stars that were never members of a multiple system. Nonetheless, single stars that were in binaries exhibit a slightly broader IMF peak and a flatter slope (by a factor of $m^{0.37}$) compared to the full single-star IMF, as shown in Extended Data Fig.~\ref{fig:mf}. This slope is consistent with the IMF of the full stellar population, including multiples. Interestingly, this difference is specific to binaries; single stars that were outer companions in higher-order multiples show no such deviation. In practice, disentangling this subtle effect from stochastic sampling and known observational biases in the IMF is challenging \citep{Kroupa2001a}, and further work is required to assess whether such signatures are detectable.
We also caution that the ejection of stars from multiples may be affected by gravitational softening, considering $\sim$70\% of ejected singles had a pericenter within two softening lengths during a multiple phase.

\section{Methods}\label{sec:meth}
\subsection{Numerical Simulations}
We use simulations from the {\sc starforge} project \citep{GrudicGuszejnov2022a,GuszejnovGrudic2022a}. {\sc starforge} is a state-of-the-art numerical framework built around the {\sc gizmo}  Lagrangian meshless finite-mass MHD code \citep{HopkinsRaives2016a,GrudicGuszejnov2021a} for the purpose of modeling the star formation process from giant molecular clouds to individual stars.

{\sc{starforge}} follows a similar approach to previous Lagrangian 3D star formation simulations \cite{klessen&burkert2000,bate+2003}, integrating the evolution of discrete mass elements according to the MHD equations. For computational efficiency, sink particles  are inserted in gravitationally collapsing regions that exceed the Jeans criterion \cite{bate+1995}, i.e., $\rho \sim 3 \times 10^{-14}$~g~cm$^{-3}$. (See \cite{grudic+2021Starforge} for the full list of sink insertion crieria.) The sink particles interact with the gas through gravity, accretion, and feedback, as determined by a sub-grid model for (proto-)star evolution \cite{offner+2009feedback,grudic+2021Starforge}. The advances achieved by {\sc starforge} are due to more detailed physics and an accurate Lagrangian approach, combined with extensive work to improve computational efficiency.

The starting point is a uniform density, spherical cloud generated by {\sc{MakeCloud}} \cite{Grudic_MakeCloud_2021}. The cloud is placed at the center of a periodic box that is 10 cloud radii across, and evolves in isolation (without any external tidal field). The initial velocity is a Gaussian random field with a power spectrum $E_k \propto k^{-2}$ \cite{ostrikerStoneGammie2001}.

The fiducial simulations (\textbf{M2e4\_mu1.3} simulations from \citep{guszejnov+2022a}) begin with an initial cloud mass of 20,000 M$_\odot$,  radius of 10 pc, and ratio of twice the kinetic to gravitational energy of $\alpha_{\rm vir}=2$. The initial magnetic field 6 $\mu$G corresponds to a mass-to-flux ratio, $\mu$, of 1.3. Stars follow a sub-grid prescription for stellar evolution that is coupled to feedback models for radiation, protostellar jet launching, stellar winds, and supernovae \citep{GrudicGuszejnov2021a}. The calculation ends when stellar feedback disperses the gas cloud, halting star formation, which occurs after $\sim$10 Myr. 

We analyze three realizations of the fiducial simulations that have different initial random seeds.(Only one of these realizations was previously published.) Together, these realizations formed a total of 6405 stars.
To test the robustness of our results, we also analyze clouds with initial magnetic fields $B_z \approx 2~\mu$G and $B_z \approx 20~\mu$G ($\mu = 4.2$ and 0.42 respectively), clouds with $\alpha_{\rm vir}=1$ and 4, and one cloud with subsolar metallicity ($Z/Z_{\odot}=0.1$). Finally, we analyzed a simulation with fiducial parameters from the newer \texttt{v1.2} {\sc{starforge}} simulations. We note that the timescale until gas expulsion, and thus the stop time of the simulation can vary with simulation parameters. Of the simulations analyzed, $\mu=0.42$ case runs for the longest physical time ($\sim 17$ Myr), while the $\alpha_{\rm vir}=1$ runs for the shortest time ($\sim 6$ Myr). 

The \texttt{v1.2} simulation analyzed here has several incremental improvements and fixes compared to those presented in e.g. \citep{GrudicGuszejnov2022a,GuszejnovGrudic2022a}, of which two are most important. First, the dust temperature is computed on-the-fly accounting for gas-grain heat transfer in addition to radiative absorption and emission, resulting in a more-realistic thermal structure in dense ($\gtrsim 10^5 \, \rm cm^{-3}$) gas. Second, the reduced speed of light has been increased from $30\,\rm km\,s^{-1}$ to $90\,\rm km \,s^{-1}$, which improves the accuracy of the radiative transfer solution in dense, optically-thick envelopes around accreting massive stars. All the simulations we analyzed have a mass resolution of $10^{-3} M_{\odot}$.

\subsection{Binary (and Multiple) Identification}
\label{sec:binID}
We use the method described in \cite{GuszejnovRaju2023a} to identify binaries and multiple systems. Briefly, at every simulation snapshot we compute the binding energy of all pairs of nearby objects, considering the fifty nearest neighbors for each star. Bound pairs are grouped into systems hierarchically, starting from the most bound. Once a system is identified, we recompute the binding energy between all pairs of objects and then proceed to the next-most-bound pair. In our analysis, systems can have no more than four stars, i.e., we do not look for hierarchies beyond quadruples. Other pairing algorithms pair objects by distance rather than binding energy \citep{bate2009}, which may affect the multiple hierarchies \citep{lee+2019}. To test the sensitivity of binary identification to the pairing algorithm, we carried out a second multiple search, pairing stars starting from the smallest semi-major axis. We found the list of binaries was very similar, though there were $\sim$5\% more with the semi-major axis pairing algorithm.

For our analysis, we only include binaries that exist for more than one snapshot and at least one binary orbital period in our analysis. More precisely, we require that 

\begin{equation}
    \Delta t \sum_{i} \frac{1}{p_{i}} \geq 1,
\end{equation}
where $\Delta t=24.7$ kyr is the time interval between snapshots, $p_{i}$ is the orbital period at snapshot $i$, and the summation is over all snapshots where the binary stars orbit each other, possibly as the inner binary of a higher order multiple. We find similar results  when replacing the right-hand-side with 10. In this case the total number of binaries is reduced by $\sim 10\%$, but the bound-from-birth fraction is consistent.

\subsection{Effects of gas on multiple identification}
\label{sec:gas}
Stars form with a bound gas halo that can affect the binding energies of stellar pairs. For example, systems that appear unbound when only the stellar masses are taken into account may actually be bound when the gas halos are considered. 

We approximately calculate the mass of the gas halo around each star particle at each snapshot as follows. First, we sort all of the gas cells by distance from the star particle. We consider a gas cell, $g$, to be bound if it satisfies three conditions. 
Firstly, 

\begin{equation}
    E_G + E_T + E_K < 0,
\end{equation}
where $E_G$ is the gravitational potential energy of the gas cell due to the star and previously identified bound gas; $E_T$ is the thermal energy of the gas; $E_K$ is the kinetic energy of the gas, star, and previously identified bound gas (in the centre-of-mass frame).  Secondly the gas must be tidally stable. Finally, the gas must be within 0.5 pc of the star particle, and there must be no closer star particle. Without the last criterion a gas cell may be identified as part of multiple halos in some cases.

For tidal stability we check that 

\begin{equation}
    \left| \underbrace{\left(\mathbf{a}_{g} - \mathbf{a}_{\rm com}\right) - \mathbf{a}_{g,{\rm halo}}}_{\mathbf{a_{t,i}}} \right| < f_t \left|\mathbf{a}_{g, {\rm halo}}\right|,
\end{equation}
where $\mathbf{a}_{g}$ is the acceleration of a gas cell  due to all the star and gas cells in the simulation, $\mathbf{a_{\rm com}}$ is the center-of-mass acceleration (of this cell, the star, and the previously identified halo particles),  $\mathbf{a}_{g,{\rm halo}}$ is the acceleration of this cell due to the star and previously identified halo particles, thus, we account for the self-gravity of the halo. Overall, the left-hand-side is the tidal acceleration of the gas cell, $a_{t,i}$. The factor $f_t$ is an adjustable parameter, which we introduce in light of results on tidal stability in Keplerian potentials.  In particular, the maximum stable binary separation can vary by a factor of two depending on the relative orientations of the internal binary orbit and the orbit of the center of mass. Since tidal forces scale as the inverse cube of separation, this corresponds to an eightfold variation in the tidal force.

In our analysis, we consider $f_t=1$ and $f_t=8$. 
We also checked $f_t=0.082$, as this approximately corresponds to a stability boundary of 0.5 Hill radii for a binary ($m_1+m_2$) in orbit around a central mass, $M$, such that $M \gg m_1 \gg m_2$. This is comparable to the minimum long-term stability boundary (for prograde binaries) from \cite{grishin+2017}. However $f_t=0.082$ produces unrealistically small gas halos, such that the final halo mass is typically smaller than the final stellar mass, as shown in Extended Data Fig.~\ref{fig:accEfficiency}.

We find that gas halos are unlikely to survive more than 1 Myr \cite{offner+2025}, and we do not include any halo mass correction for stars older than 1 Myr for computational efficiency. We find that the median size of our halos is of order 1000 au, which is within a factor of few of the Jeans' length, as shown in Supplementary Fig.~\ref{fig:haloSize}. For each halo, we estimate the size as $\left(a_1 a_2 a_3\right)^{1/3}$, where $a_1$, $a_2$, and $a_3$ are the sizes of the principle axes. The Jeans' length is estimated using the (mass-weighted) mean density and sound speed of the halo. The maximum halo mass is typically $\sim$5 times larger than the final star mass.

We then identify multiples, adding the gas halo mass to the star mass for each star and snapshot. We artificially place each gas halo cell at the location of the corresponding star. In Supplementary Fig.~\ref{fig:energySpace} we demonstrate that `collapsing' the halo in this way does not qualitatively change our results (e.g. Fig.~\ref{fig:energy}).

\section{Data Availability}
The full snapshots, containing gas and sink data from the underlying {\sc staforge} simulations are available upon request.

\section{Code Availability}
The scripts used to compute halos, identify multiples and generate figures are published as a Code Ocean capsule at https://doi.org/10.24433/CO.6648239.v1. We made use of ChatGPT for code refactoring.

This project made use of the open source package pytreegrav (\url{https://github.com/mikegrudic/pytreegrav}) for the calculation of tidal forces. The open source package Meshoid (\url{https://github.com/mikegrudic/meshoid}) was used for the visualization in Fig. 1.
{\sc{Starforge}} uses a numerical framework implemented in the Gizmo code. The public version of the Gizmo code, which includes self-gravity, MHD, radiation transfer and various other physics modules is available at \url{https://bitbucket.org/phopkins/gizmo-public/src/master/}. 
\backmatter

\bmhead{Acknowledgements}
We thank the anonymous referees for constructive feedback that dramatically improved the quality of the paper. We thank Juan Farias for helpful comments and discussions. A.G. was supported at the Technion by a Zuckerman Fellowship. A.G., S.O., K.K. acknowledge support from NSF AAG 2407522. S.O. also acknowledges support from a Peter O'Donnell Research Fellowship and a Donald Harrington Faculty Fellowship.

The authors acknowledge the Texas Advanced Computing Center (TACC) at The University of Texas at Austin for providing computational resources that have contributed to the research results reported within this paper. URL: http://www.tacc.utexas.edu

\bmhead{Author contributions}
A.G. developed code for multiple and gas halo identification, with help from D.G, and carried out the analysis. M.G. ran the {\sc Starforge} simulation models.
S.O. provided expertise related to the {\sc Starforge} simulations and facilitated the analysis.
A.G, S.O, K.K, and H.B.P. contributed to the interpretation and discussion of the results and writing and editing of the paper.

\bmhead{Competing interests}
The authors declare no competing interests.

\bmhead{Materials \& Correspondence}
 Correspondence and material requests should be addressed to A.G.


\clearpage

\section*{Extended data}
\begin{extendedtable}[h]
\caption{Summary of binary statistics for simulations with three different virial parameters (and a fixed mass-to-flux ratio, $\mu=4.2$). The last two columns show data for a low metallicity ($Z/Z_{\odot}=0.1$) simulation. We show data for individual simulations with different initial random seeds on the bottom of each cell, and stacked data at the top in bold.}\label{tab:binStatsVir}
\begin{tabular}{|l|c|c|c|c|c|c|c|c|}
\hline
 & \multicolumn{2}{c|}{\texttt{alpha2}} & \multicolumn{2}{c|}{\texttt{alpha1}} & \multicolumn{2}{c|}{\texttt{alpha4}} & \multicolumn{2}{c|}{\texttt{alpha2,z0.1}} \\
 & \multicolumn{2}{c|}{seeds=3} & \multicolumn{2}{c|}{seeds=2} & \multicolumn{2}{c|}{seeds=2} & \multicolumn{2}{c|}{seeds=1} \\
\hline
 & Count & Frac & Count & Frac & Count & Frac & Count & Frac \\
\hline
\multicolumn{9}{|c|}{\textbf{All}}\\
\hline
Total &
\begin{tabular}{@{}c@{\hskip 1pt}c@{\hskip 1pt}c@{}}
 & \textbf{1792} & \\
{\footnotesize 545} & {\footnotesize 600} & {\footnotesize 647}
\end{tabular} 
& 
\begin{tabular}{@{}c@{\hskip 1pt}c@{\hskip 1pt}c@{}}
 & 1.00 & \\
 & & 
\end{tabular} 
&
\begin{tabular}{@{}c@{\hskip 1pt}c@{}}
   \multicolumn{2}{c}{\textbf{1370}} \\
{\footnotesize 551} & {\footnotesize 819}
\end{tabular} 
& 
\begin{tabular}{@{}c@{\hskip 1pt}c@{}}
   \multicolumn{2}{c}{1.00} \\
 & 
\end{tabular} 
&
\begin{tabular}{@{}c@{\hskip 1pt}c@{}}
   \multicolumn{2}{c}{\textbf{994}} \\
{\footnotesize 544} & {\footnotesize 450}
\end{tabular} 
&
\begin{tabular}{@{}c@{\hskip 1pt}c@{}}
   \multicolumn{2}{c}{1.00} \\
 & 
\end{tabular} 
&
\begin{tabular}{@{}c@{\hskip 1pt}c@{}}
   \multicolumn{2}{c}{\textbf{341}} \\
 &
\end{tabular} 
&
\begin{tabular}{@{}c@{\hskip 1pt}c@{}}
   \multicolumn{2}{c}{1.00} \\
   &
\end{tabular}\\
\hline
Bound &
\begin{tabular}{@{}c@{\hskip 1pt}c@{\hskip 1pt}c@{}}
 & \textbf{1301} & \\
{\footnotesize 405} & {\footnotesize 436} & {\footnotesize 460}
\end{tabular} 
& 
\begin{tabular}{@{}c@{\hskip 1pt}c@{\hskip 1pt}c@{}}
 & \textbf{0.73} & \\
{\footnotesize 0.74} & {\footnotesize 0.73} & {\footnotesize 0.71}
\end{tabular} 
&
\begin{tabular}{@{}c@{\hskip 1pt}c@{}}
   \multicolumn{2}{c}{\textbf{971}} \\
{\footnotesize 407} & {\footnotesize 564}
\end{tabular} 
&
\begin{tabular}{@{}c@{\hskip 1pt}c@{}}
   \multicolumn{2}{c}{\textbf{0.71}} \\
{\footnotesize 0.74} & {\footnotesize 0.69}
\end{tabular} 
&
\begin{tabular}{@{}c@{\hskip 1pt}c@{}}
   \multicolumn{2}{c}{\textbf{685}} \\
{\footnotesize 372} & {\footnotesize 313}
\end{tabular} 
&
\begin{tabular}{@{}c@{\hskip 1pt}c@{}}
   \multicolumn{2}{c}{\textbf{0.69}} \\
{\footnotesize 0.68} & {\footnotesize 0.70}
\end{tabular} 
&
\begin{tabular}{@{}c@{\hskip 1pt}c@{}}
   \multicolumn{2}{c}{\textbf{244}} \\
 &
\end{tabular} 
&
\begin{tabular}{@{}c@{\hskip 1pt}c@{}}
   \multicolumn{2}{c}{\textbf{0.72}} \\
   &
\end{tabular}\\
\hline
\multicolumn{9}{|c|}{\textbf{Surviving}}\\
\hline
Total &
\begin{tabular}{@{}c@{\hskip 1pt}c@{\hskip 1pt}c@{}}
 & \textbf{938} & \\
{\footnotesize 292} & {\footnotesize 318} & {\footnotesize 328}
\end{tabular} 
& 
\begin{tabular}{@{}c@{\hskip 1pt}c@{\hskip 1pt}c@{}}
 & 1.00 & \\
 & & 
\end{tabular} 
&
\begin{tabular}{@{}c@{\hskip 1pt}c@{}}
   \multicolumn{2}{c}{\textbf{698}} \\
{\footnotesize 294} & {\footnotesize 404}
\end{tabular} 
& 
\begin{tabular}{@{}c@{\hskip 1pt}c@{\hskip 1pt}c@{}}
 & 1.00 & \\
 & & 
\end{tabular} 
&
\begin{tabular}{@{}c@{\hskip 1pt}c@{}}
   \multicolumn{2}{c}{\textbf{492}} \\
{\footnotesize 257} & {\footnotesize 235}
\end{tabular} 
& 
\begin{tabular}{@{}c@{\hskip 1pt}c@{\hskip 1pt}c@{}}
 & 1.00 & \\
 & & 
\end{tabular} 
&
\begin{tabular}{@{}c@{\hskip 1pt}c@{}}
   \multicolumn{2}{c}{\textbf{174}} \\
 &
\end{tabular} 
&
\begin{tabular}{@{}c@{\hskip 1pt}c@{}}
   \multicolumn{2}{c}{1.00} \\
   &
\end{tabular}\\
\hline
Bound &
\begin{tabular}{@{}c@{\hskip 1pt}c@{\hskip 1pt}c@{}}
 & \textbf{670} & \\
{\footnotesize 212} & {\footnotesize 220} & {\footnotesize 238}
\end{tabular} 
& 
\begin{tabular}{@{}c@{\hskip 1pt}c@{\hskip 1pt}c@{}}
 & \textbf{0.71} & \\
{\footnotesize 0.73} & {\footnotesize 0.69} & {\footnotesize 0.73}
\end{tabular} 
&
\begin{tabular}{@{}c@{\hskip 1pt}c@{}}
   \multicolumn{2}{c}{\textbf{492}} \\
{\footnotesize 218} & {\footnotesize 274}
\end{tabular} 
&
\begin{tabular}{@{}c@{\hskip 1pt}c@{}}
   \multicolumn{2}{c}{\textbf{0.70}} \\
{\footnotesize 0.74} & {\footnotesize 0.68}
\end{tabular} 
&
\begin{tabular}{@{}c@{\hskip 1pt}c@{}}
   \multicolumn{2}{c}{\textbf{326}} \\
{\footnotesize 158} & {\footnotesize 168}
\end{tabular} 
&
\begin{tabular}{@{}c@{\hskip 1pt}c@{}}
   \multicolumn{2}{c}{\textbf{0.66}} \\
{\footnotesize 0.65} & {\footnotesize 0.67}
\end{tabular} 
&
\begin{tabular}{@{}c@{\hskip 1pt}c@{}}
   \multicolumn{2}{c}{\textbf{123}} \\
 &
\end{tabular} 
&
\begin{tabular}{@{}c@{\hskip 1pt}c@{}}
   \multicolumn{2}{c}{\textbf{0.71}} \\
   &
\end{tabular}\\
\hline
\end{tabular}
\end{extendedtable}

\clearpage

\clearpage
\begin{extendedfigure}[h]
    \centering
    \includegraphics[width=\textwidth]{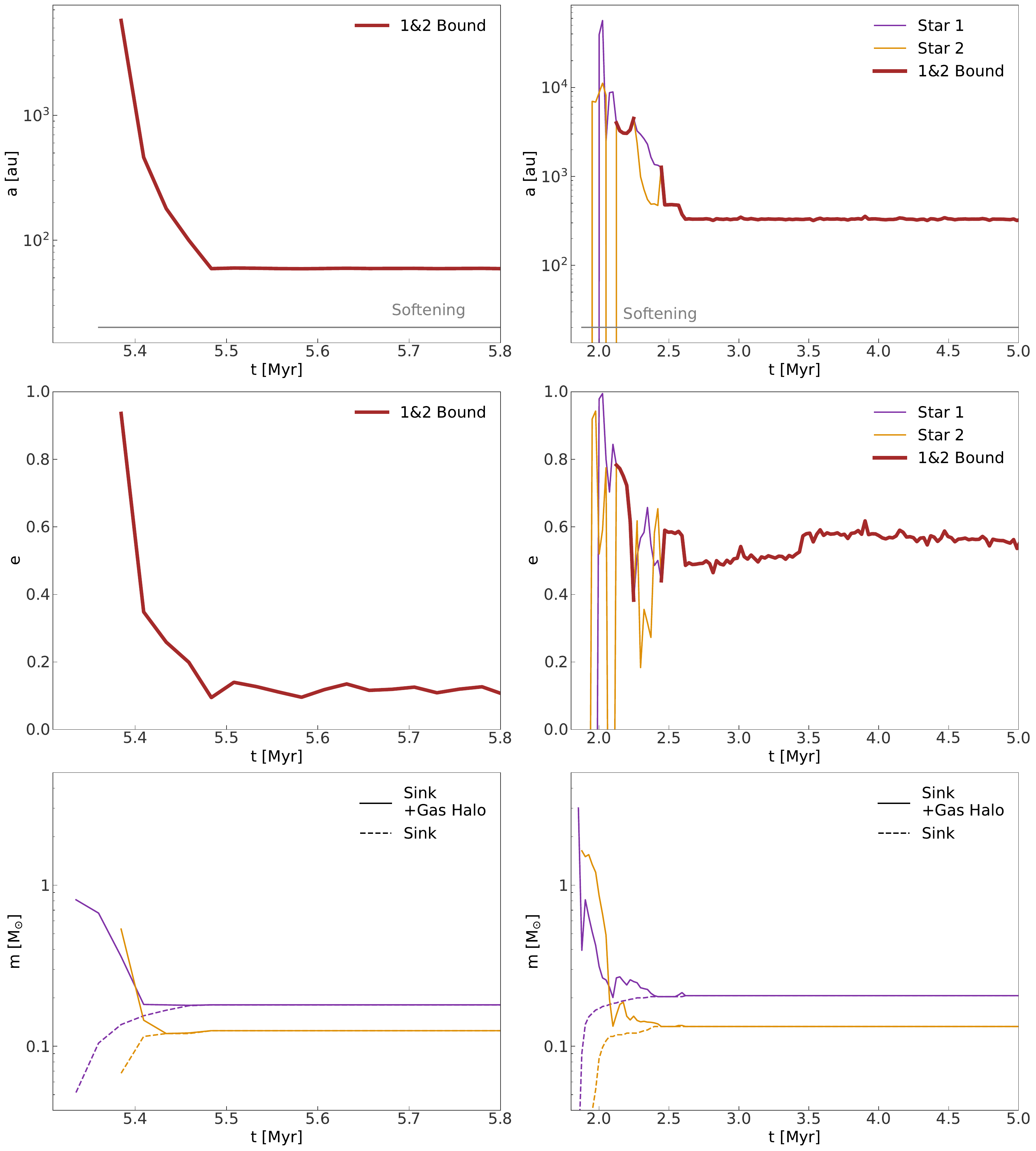}
    \caption{Early evolution of the semi-major axes (top), eccentricities (middle) for example binaries. When the binary is bound, the eccentricity and semi-major axis are a thick, brown line. If the stars' are bound to other objects their orbital elements are thin orange or purple lines. For reference, we show the stars' gravitational softening length as a horizontal gray line in the top panels. The bottom panels show the mass evolution for these binaries. The solid lines show the mass of each star and its halo, while the dashed lines show the masses of the stars alone.}\label{fig:evolve}
\end{extendedfigure}
\clearpage

\clearpage

\clearpage
 \begin{extendedfigure}[h]
    \centering
    \includegraphics[width=\linewidth]{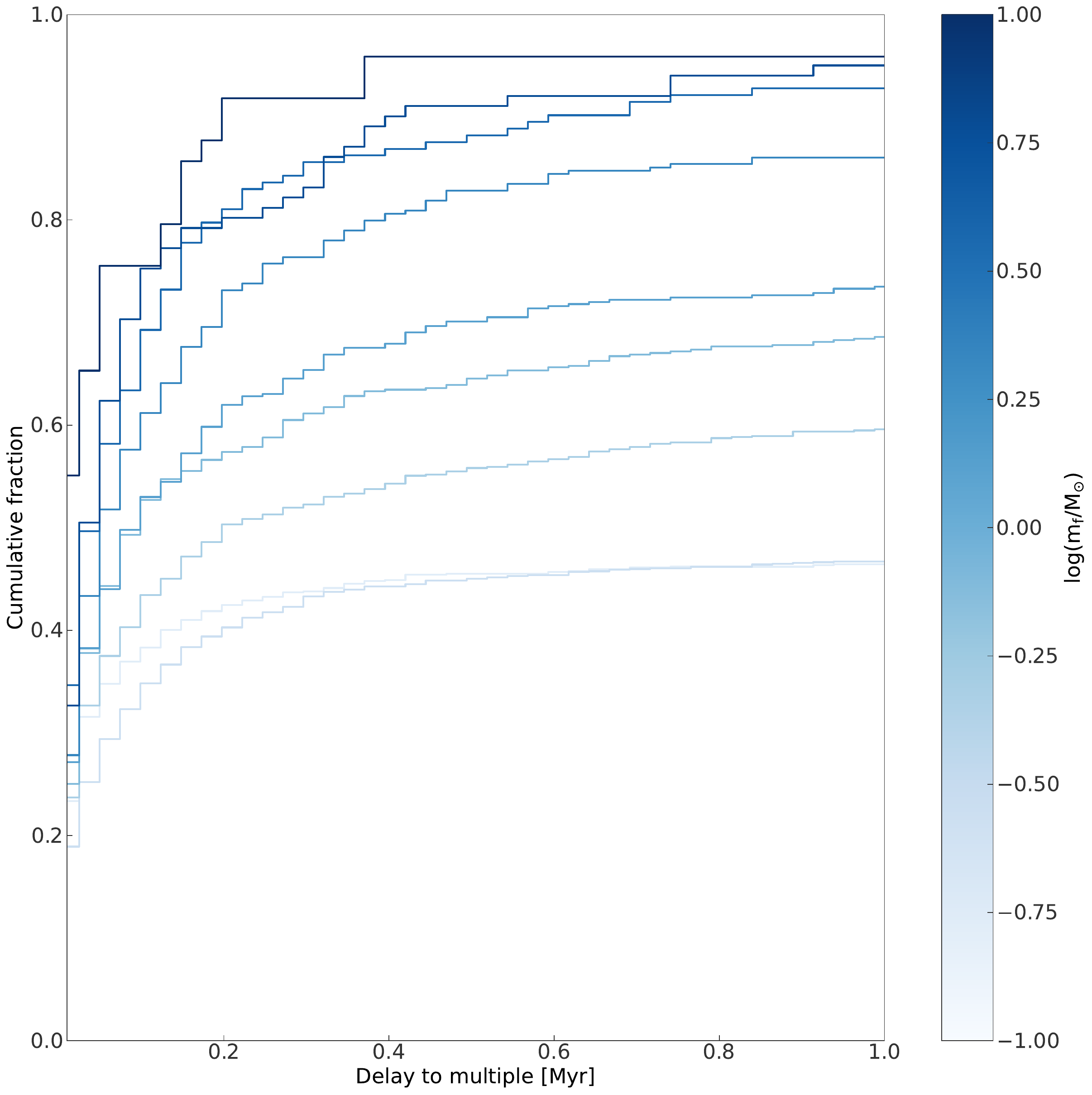}
    \caption{Cumulative distribution of the delay from formation of stars to their first identification in a (persistent) multiple system (binary, triple, or quadruple). Each coloured line corresponds to 1 of 10 logarithmically spaced mass bins, where stars are binned by their final masses. Persistent multiples are those that survive for more than one snapshot and at least one period. Note that many low mass stars are never in multiples, so the maximum cumulative fraction remains below one.}
    \label{fig:delayMult}
\end{extendedfigure}

\clearpage
 \begin{extendedfigure}[h]
    \centering
    \includegraphics[width=\linewidth]{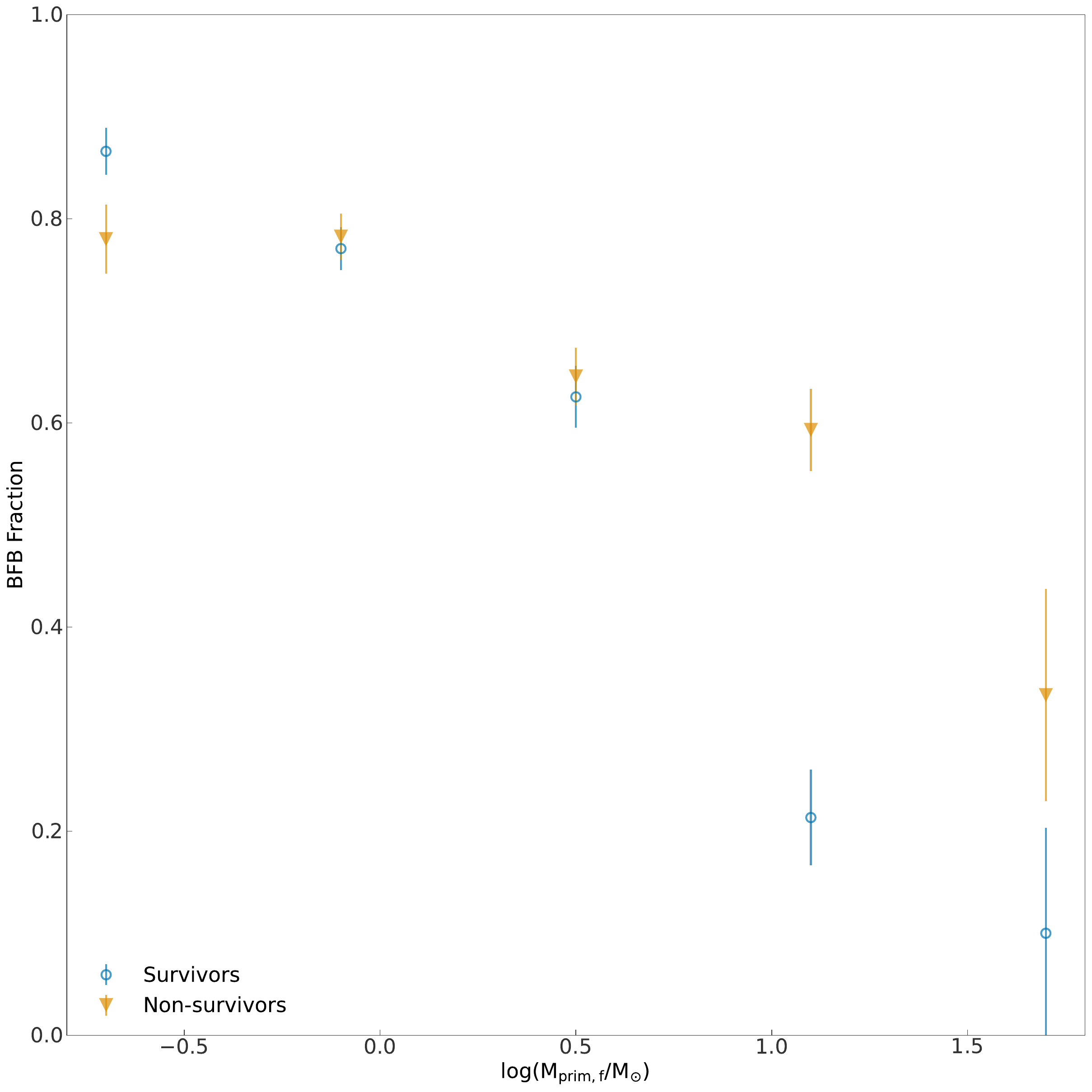}
    \caption{\label{fig:bfbMass} Bound from birth fraction versus mass for surviving (blue, open circles) and non-surviving binaries (orange, filled triangles). The mass is the maximum of the two stars' final masses. The error bars show the standard deviation, from the Bayesian formula $\sqrt{\frac{(N - k + 1) (k + 1)}{(N+ 3) (N + 2)^2}}$ \cite{guszejnov+2023}, where $k$ is the number of bound binaries and $N$ is the total number of binaries in the mass bin. From left to right $k$ is 194, 306, 157,  16, 1 for the survivors and 117, 259, 182,  86,  6 for the non-survivors. From left to right $N$ is 224, 397, 251, 75,  10 for the survivors and 150, 331, 282, 145,  18 for the non-survivors.  While high mass stars tend to form in multiple systems, many of these are disrupted by exchanges, leading to a decline in the overall bound-from-birth fraction with mass. 
    }
\end{extendedfigure}

\clearpage
\begin{extendedfigure}[h]
    \includegraphics[width=\linewidth]{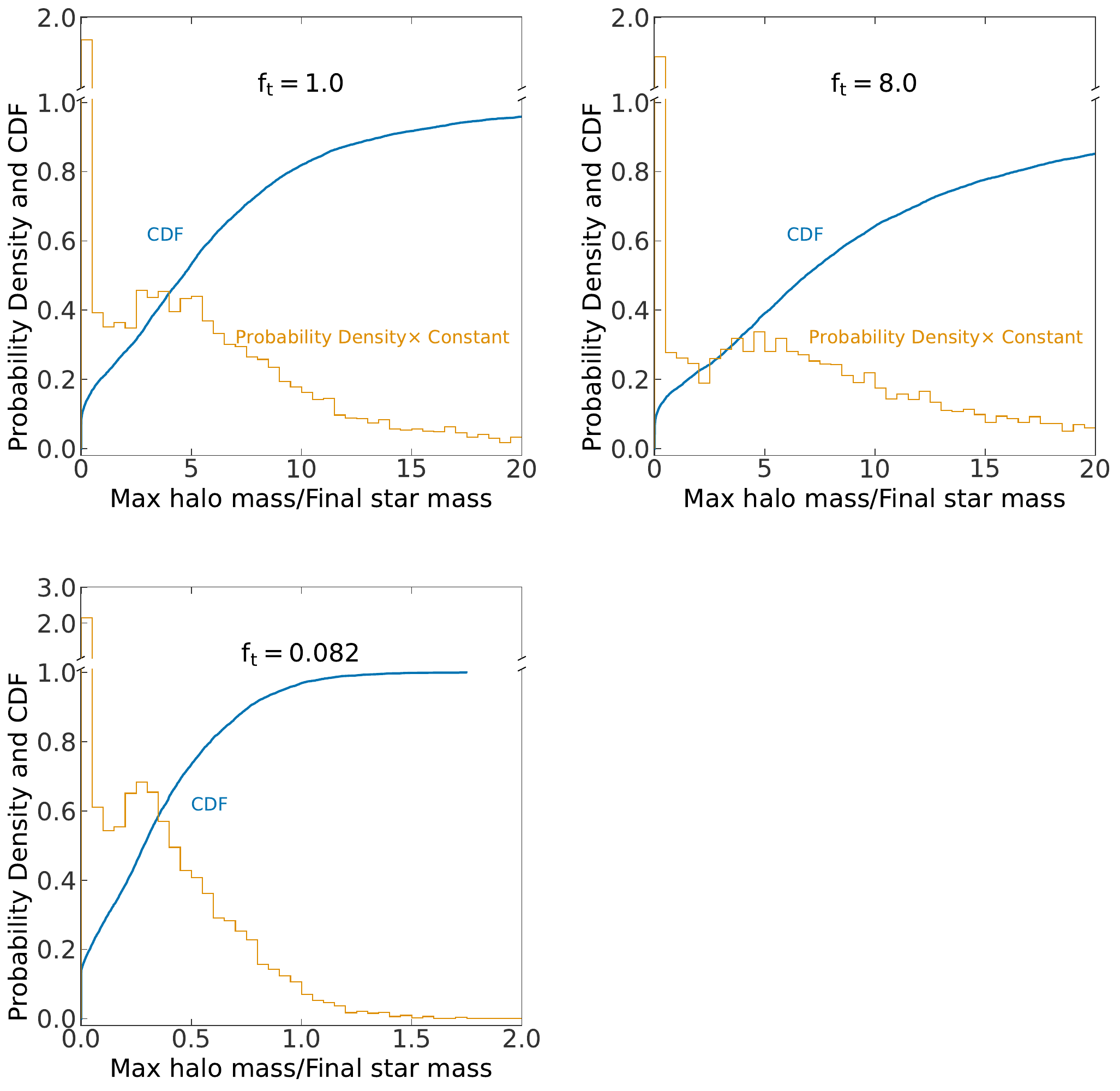}
    \caption{\label{fig:accEfficiency} PDF (orange) and CDF (blue) of the ratio between the maximum gas halo mass and the final stellar mass for all of the stars in the simulation for $f_t=1$ (top left), $f_t=8$ (top right), and $f_t=0.082$ (bottom). Note the different axis ranges in the bottom panel.}
\end{extendedfigure}

\clearpage
\begin{extendedfigure}[h]
    \includegraphics[width=\linewidth]{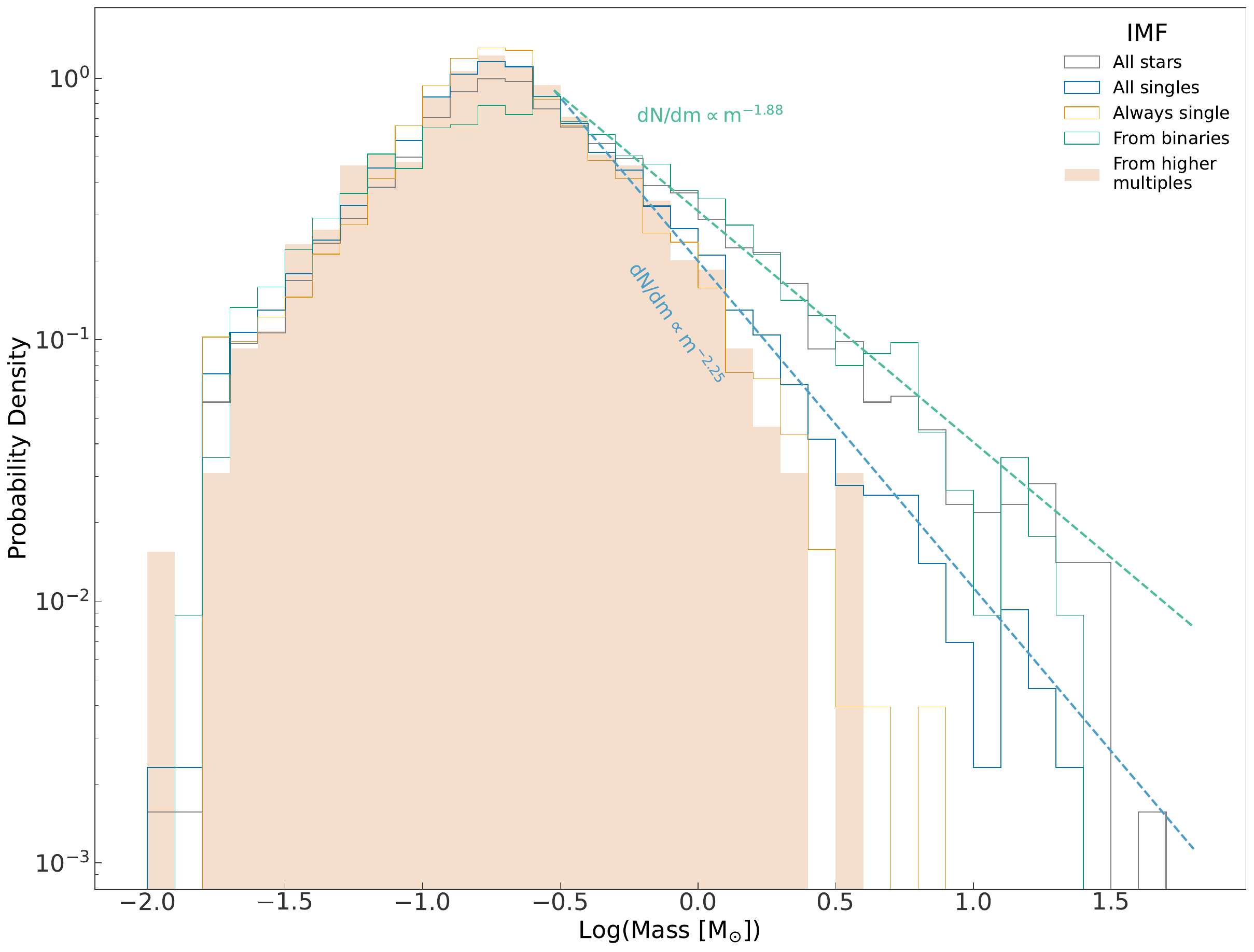}
    \caption{\label{fig:mf} Stellar initial mass function (IMF). The blue histogram shows the overall IMF, while the green histogram shows the IMF of singles that were in binaries at some point. The latter has a flatter slope above 0.3 $M_{\odot}$, as indicated by the power-law fits (dashed blue and green lines). The flatter slope is comparable to that of the IMF of all stars in the simulation (gray histogram). The dash-dotted orange histogram shows the IMF of singles that were in higher multiples, but not in binaries (indicating they are outer companions). The solid orange histogram shows the IMF of stars that were never in multiple systems.}
\end{extendedfigure}

\section*{Supplementary information}
\pagenumbering{gobble}
\begin{supfigure}[H]
\centering
\begin{minipage}[b]{0.48\linewidth}
\centering
\includegraphics[width=\linewidth]{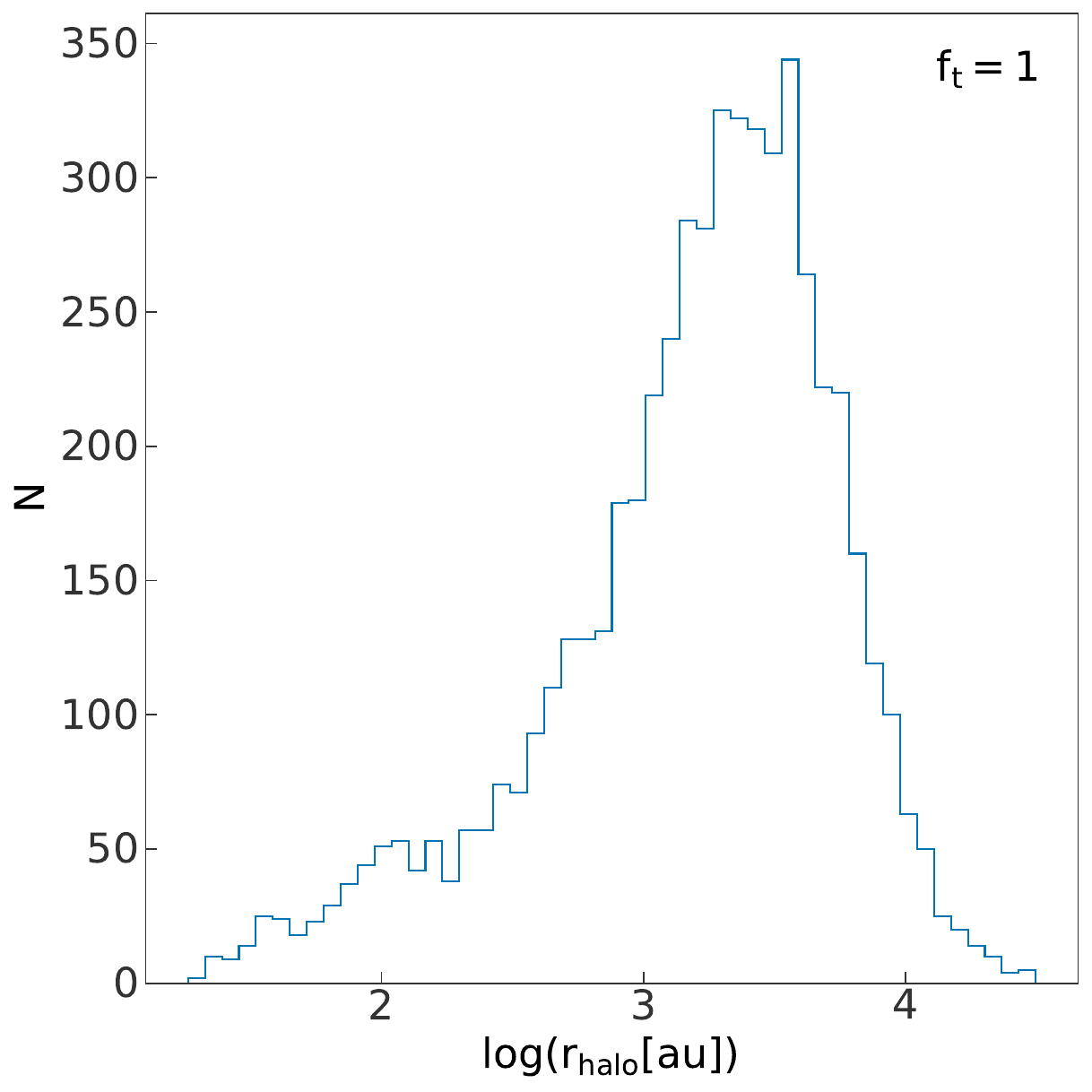}
\end{minipage}
\hfill
\begin{minipage}[b]{0.48\linewidth}
\centering
\includegraphics[width=\linewidth]{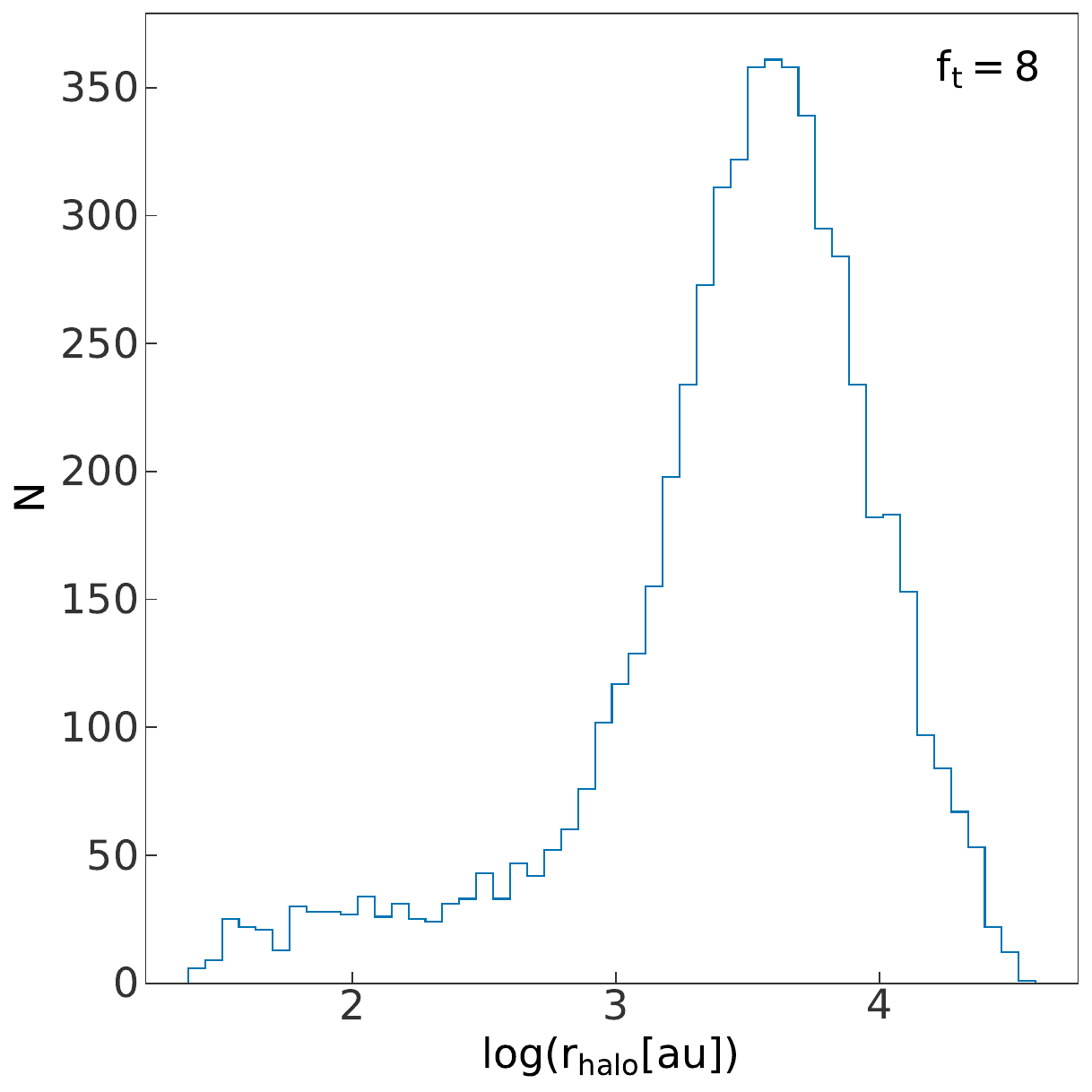}
\end{minipage}

\vspace{2mm} 

\begin{minipage}[b]{0.48\linewidth}
\centering
\includegraphics[width=\linewidth]{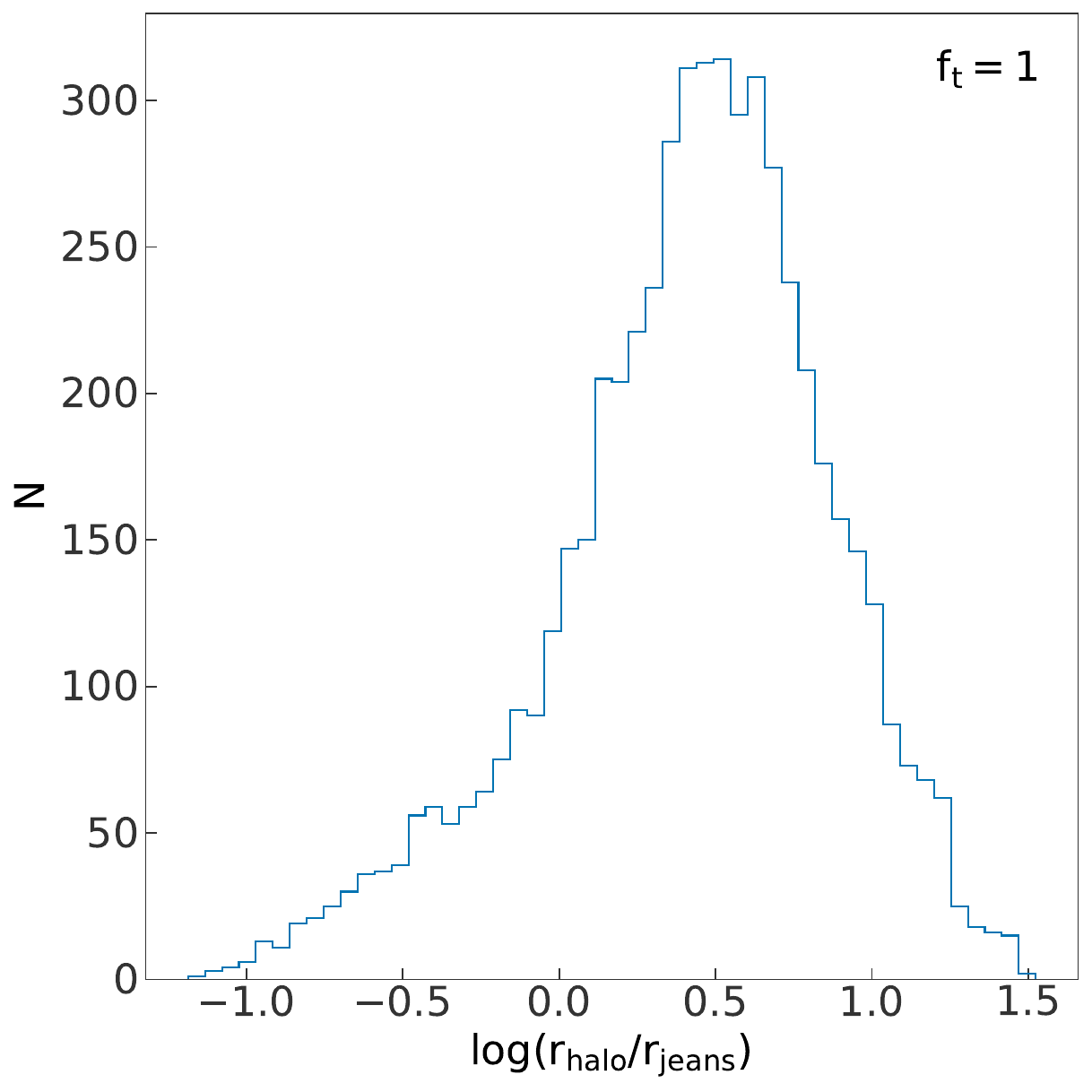}
\end{minipage}
\hfill
\begin{minipage}[b]{0.48\linewidth}
\centering
\includegraphics[width=\linewidth]{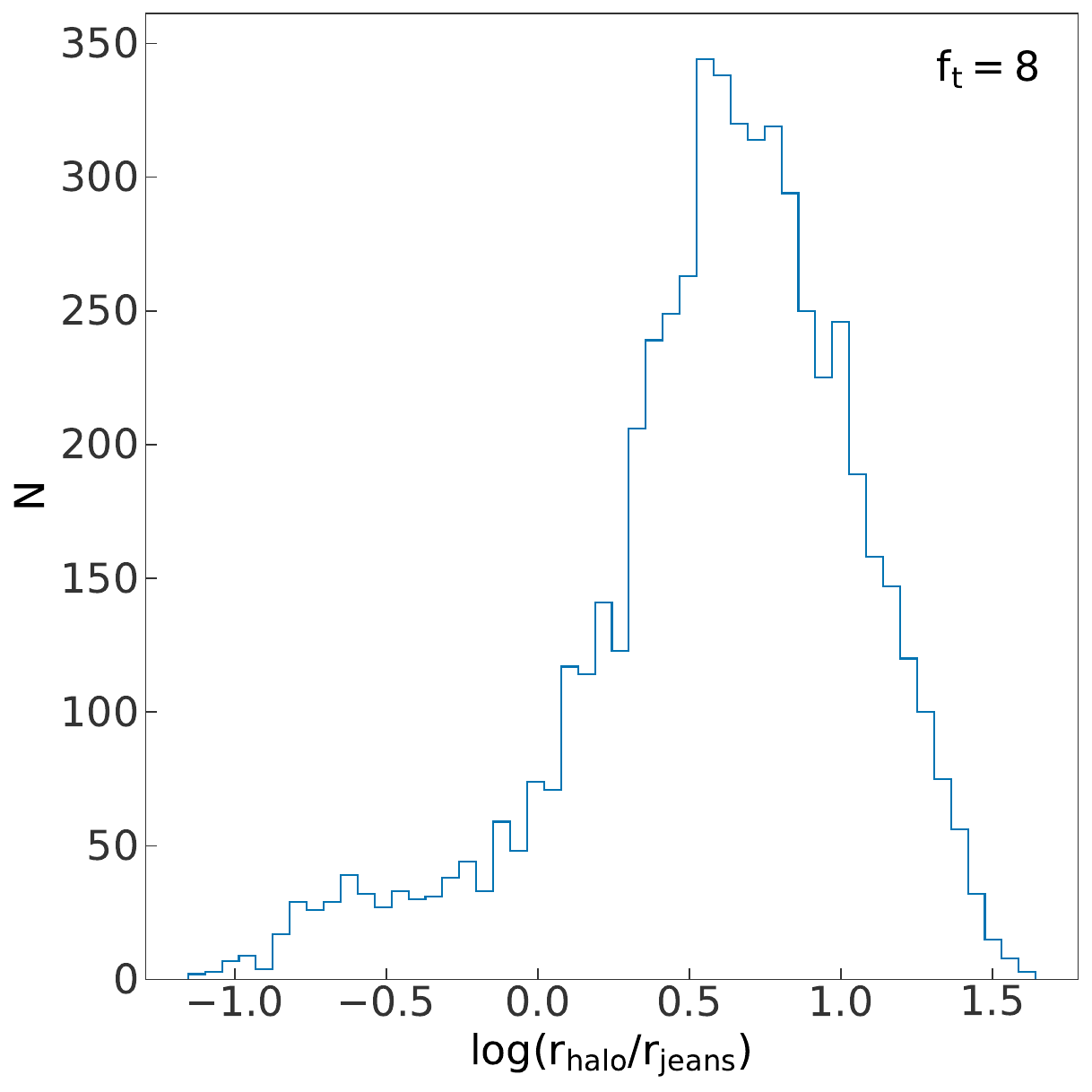}
\end{minipage}

\vspace{2mm}

\begin{minipage}[b]{0.48\linewidth}
\centering
\includegraphics[width=\linewidth]{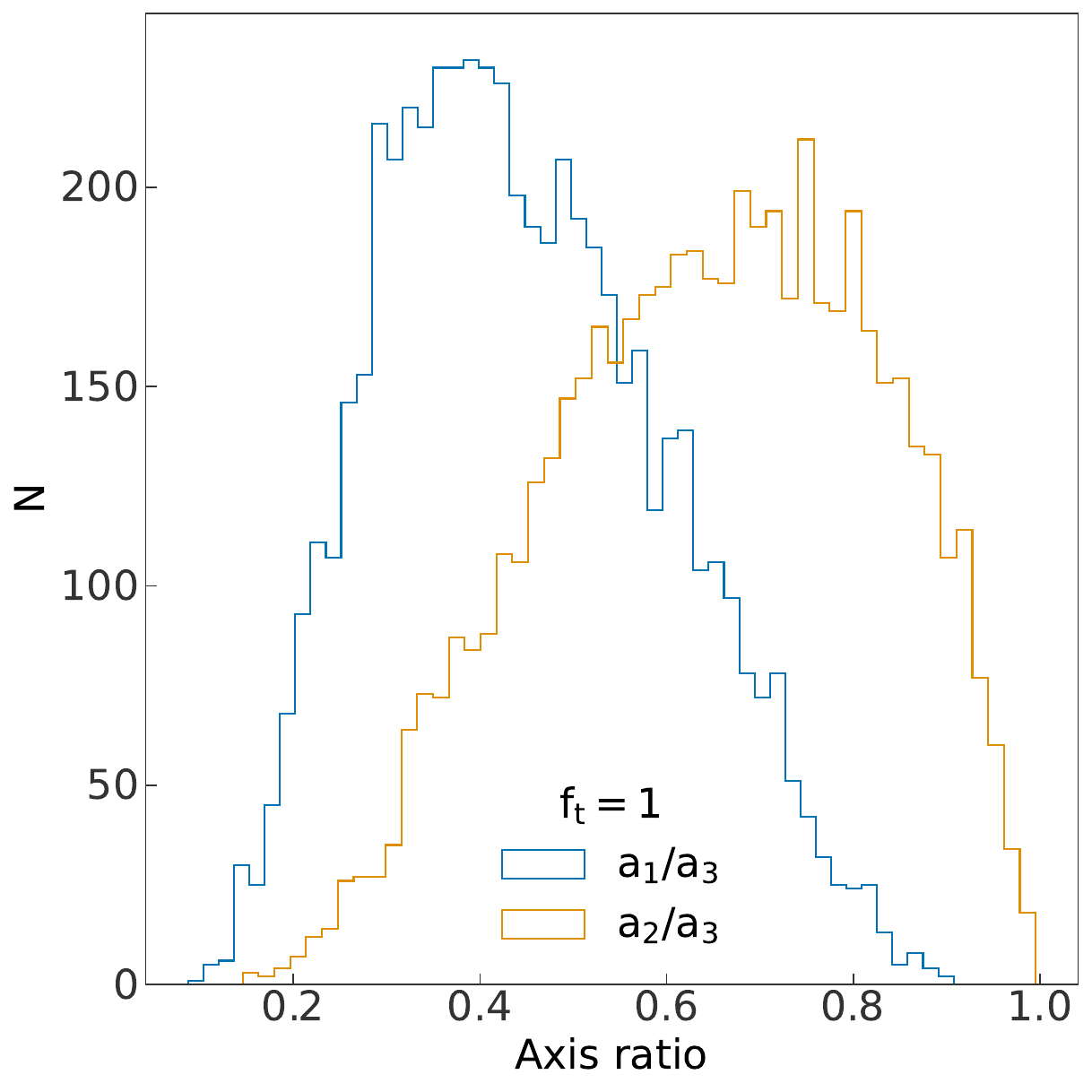}
\end{minipage}
\hfill
\begin{minipage}[b]{0.48\linewidth}
\centering
\includegraphics[width=\linewidth]{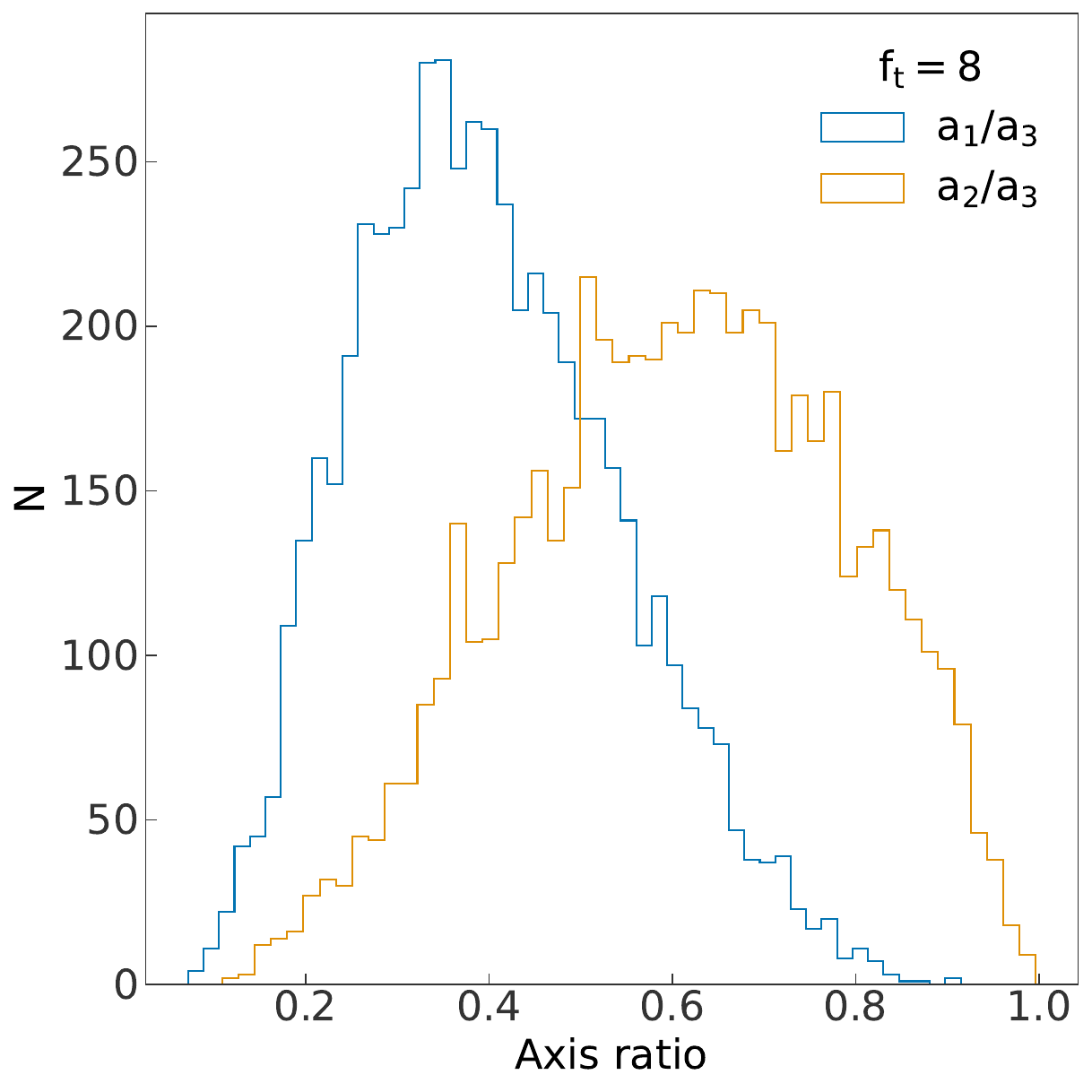}
\end{minipage}
\caption{\label{fig:haloSize} The distribution of the halo sizes at sink formation for $f_t=1$ (left) and $f_t=8$ (right). The middle row show the ratio of halo size to Jeans' length (note the different scale for the x-axis). The halo size is estimated as $(a_1 a_2 a_3)^{1/3}$, where $a_1$, $a_2$, and $a_3$ are the (sorted) lengths of the halo's principal axes. The Jean's length is calculated using the (mass-weighted) average density and sound speed of the halo. The bottom row shows the axis ratios of the halos.}
\end{supfigure}
\pagenumbering{gobble}

\clearpage

 \begin{supfigure}
    \centering
    \includegraphics[width=\textwidth]{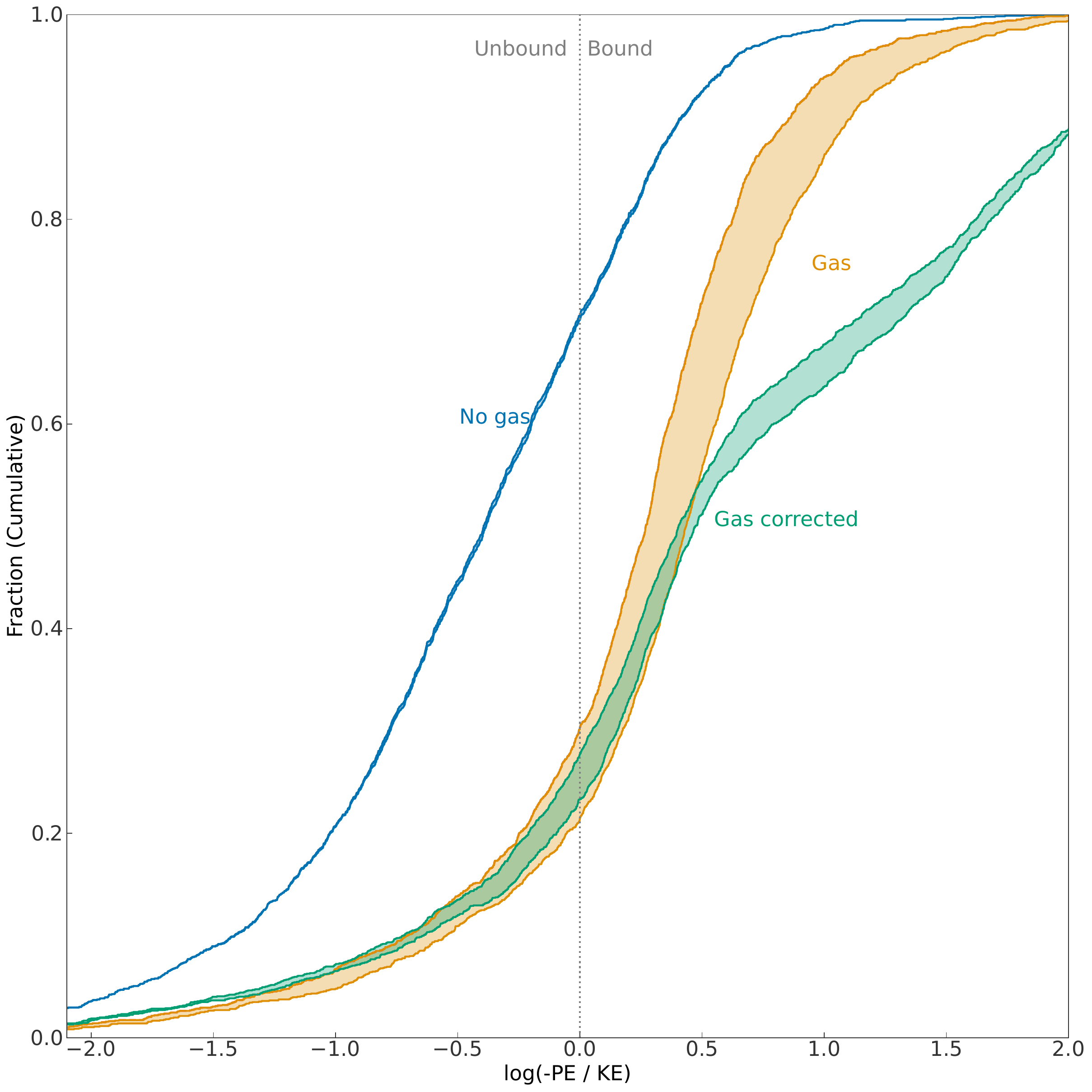}
    \caption{Binding energy of all binary pairs at
the IST, normalized to the kinetic energy (in the center of mass frame). Generally, when this ratio
is greater than one, the binary is bound from birth. Conversely, when the ratio is less than one
there is a delay between star and binary formation.  The number of pairs in the unbound
group is significantly reduced if the gas halo mass is included in the binary binding energy, as shown
by the difference between the blue (no gas) and the orange and green histograms (gas). In the former, the gas halos are included in the potential energy by adding their mass to that of each star. In the latter, we calculate the potential energy of the secondary star in each binary, due to the primary and its gas halo (with the correct particle positions). The green line shows the ratio of this potential energy to the kinetic energy:
$KE = \frac{1}{2} \left[m_2 v_2^2 + (m_1 + m_{\rm gas,1}) v_1^2\right]$,
 where $m_2$, $m_1$, and $m_{\rm gas,1}$ are the masses of the secondary, primary, and the primary gas halo, $v_2$ is the velocity of the secondary, and $v_1$ is the velocity of the center-of-mass of the primary and its gas halo. All velocities are evaluated in the center-of-mass frame.  The maximum fraction of unbound binaries remains $\sim$25\%. 
}
    \label{fig:energySpace}
\end{supfigure}

\end{document}